\documentclass{article}

\usepackage{graphics,amsmath,graphicx,bm,hyperref}
\usepackage{color}
\usepackage{grffile}
\begin{document}
\title{Thermodynamic properties of the trigonometric Rosen-Morse potential and applications to a quantum gas of mesons}

\author{
Aram Bahroz Brzo\footnote{aram.brzo@univsul.edu.iq}\\
Physics Department (College of Education)\\ University of Sulaimani\\
New camp - Tasluja - Street 1- Zone 501 Sulaimania\\ As Sulaymaniyah, Iraq\\
\newline \\
David Alvarez-Castillo\footnote{alvarez@theor.jinr.ru}\\
 H.Niewodnicza\'nski Institute of Nuclear Physics, PAS\\
152 Ul. Radzikowskiego, Cracow, 31-342, Poland\\
Bogoliubov Laboratory of Theoretical Physics, JINR\\
6 Joliot-Curie St, Dubna 141980, Russian Federation\\
       }




\date{}

\maketitle
\begin{abstract}
In this study we work out thermodynamic functions for a quantum gas of mesons described as color-electric charge dipoles. 
They refer to a particular parametrization of the trigonometric Rosen-Morse potential which allows to transform it to
a perturbation of free quantum motion on the
three-dimensional hyper-sphere, $S^{3}$, a manifold that can
host only charge-neutral systems, the charge dipoles being the configuration of the minimal number of constituents.  To the amount charge
neutrality manifests itself as an important aspect of the color confinement 
in the theory of strong interaction, the Quantum Chromodynamics, we expect our
findings to be of interest to the evaluation of temperature phenomena in
the physics of hadrons and in particular in a quantum gas of color charge dipoles 
as are the mesons. The results are illustrated for $f_0$ and $J/\psi$ mesons.\\

{keywords: Trigonometric Rosen Morse potential; partition function; mesons.}
\end{abstract}


\section{Introduction}
\label{intro}

The study of the thermodynamical properties of any system begins with
the calculation of its partition function. All thermodynamical or
statistical properties can be obtained from the partition function since
it contains all the information about the given system \cite{Hoover:2012csm}.
 A basic example is provided by the ideal quantum gas of di-atomic molecules, whose properties have been extensively studied in atomic, and molecular physics \cite{Baron:2007trz,Ikot:2018chu,Ikot:2018chu:2015dio,Drellishak:1965aes}. 
Di-atomic molecules
are described be it as rigid or elastic dumbbells with the dumbbells
tracing trajectories placed on a two-dimensional space of constant
curvature, the sphere immersed into a flat three-dimensional space. 
At the level of the one-di\-men\-si\-on\-al  Schr\"odinger equation, the partition function relevant for this case is the one determined by the  trigonometric Scarf potential. Besides the textbook example of the harmonic oscillator, partition functions are known for several exactly solvable potentials, among them the hyperbolic 
Scarf \cite{Pratiwi:2017}, and the trigonometric P\"oschl-Teller potentials \cite{Edet:2019}. 

A specifically interesting case occurs when the embedding space is by
itself of constant curvature and takes the shape of the
three-dimensional sphere, $S^{3}$. Such a situation can happen in
cases of eigenvalue problems based upon interactions which are most
favorably solved in hyper-spherical coordinates. A prominent example for
this is provided by the trigonometric Rosen-Morse potential \cite{Cooper:1994eh}
given by,
\begin{equation}
\label{1}
V_{tRM}^{\left( {\bar a},b \right)} \left( \chi \right) = \frac{\hbar^{2}\ c^{2}\ }{2M\ c^{2}\ R^{2}}\left[\frac{{\bar a}\left( {\bar a} - 1 \right)}{\sin^{2}(\chi)} - 2b\cot\left( \chi \right) \right],~\chi  = \frac{r}{R} ,  
\end{equation}
\begin{equation}
0 \leq \chi \leq \pi, \nonumber
\end{equation}
where $\chi$ is an angular variable, ${\bar a}$ is a parameter, $r$ is a relative
distance in flat space, $R$ is the matching length parameter. The one-dimensional
Schr\"odinger equation with this potential,
\begin{equation}
\label{2}
\begin{split}
\frac{\hbar^{2}\ c^{2}\ }{2M\ c^{2}\ R^{2}}\left( - \frac{d^{2}\ }{d\chi^{2}} + V_{tRM}^{\left( {\bar a},b \right)}\left( \chi \right) \right)U(\chi) =  \\
\frac{\hbar^{2}\ c^{2}\ }{2M\ c^{2}\ R^{2}}\left[\left( n + {\bar a} \right)^{2} - \frac{b^{2}}{\left( n + {\bar a} \right)^{2}}\  \right] U(\chi),
\end{split}
\end{equation}
where $n$ stands for the node number of the wave function, has a
remarkable property. Namely, upon the variable and parameter changes
according to \cite{Kalnins:2002qc},
\begin{equation}
\label{3}
\Phi(\chi) = \frac{U\left( \chi \right)}{\sin\left( \chi \right)},\,\,  {\bar a} \rightarrow l + 1,\,\,  l = 0,\ 1,\ 2,
\end{equation}
i.e. for integer values of the ${\bar a}$ parameters, the equation (\ref{2}) is transformed to
\begin{eqnarray}
\left( - \frac{\hbar^{2}c^{2}}{2M c^{2}}\mathrm{\Delta}_{S^{3}}\left( \chi \right) - \ \frac{\hbar^{2}c^{2}}{2M c^{2}R^{2}}2b\cot{\left( \chi \right) -}\frac{\hbar^{2}c^{2}}{2M c^{2}R^{2}} \right)\Phi\left( \chi \right) &&\nonumber\\
= \frac{\hbar^{2}c^{2}}{2M c^{2}R^{2}}\ \left( \ k(k + 2) - \frac{b^{2}\ }{\left( k + 1 \right)^{2}\ } \right)\Phi\left( \chi \right), &&\nonumber\\
\label{4}
\end{eqnarray}
with
\begin{eqnarray}
 - \mathrm{\Delta}_{S^{3}}\left( \chi \right) &=&
 \ \frac{- 1}{R^{2}\sin^{2}\left( \chi \right)\ }\ \frac{\partial}{\partial\chi}\ \sin^{2}{\left( \chi \right)\frac{\partial}{\partial\chi} + \frac{l(l + 1)}{ R^2 \sin^{2}\left( \chi \right)}} \nonumber\\
&=& \frac{ 1}{R^{2}}K^{2}(\chi),\quad k = n + l.
\label{5}
\end{eqnarray}
Here $\mathrm{\Delta}_{S^{3}}(\chi)$ stands for the part of the Laplace-Beltrami operator on the three-dimensional hyper-sphere, $S^{3}$, of hyper-radius $R$,  
which depends on the second polar angle, $\chi$, while
$K^{2}(\chi)$ denotes the squared operator of the four-dimensional
angular momentum in the same variable. 
The second polar angle parametrizing the hyper-sphere is now given by $\chi=\stackrel{\frown}{r}/R$
where $\stackrel{\frown}{r}$ is the arc of the particle's path read off from the North-pole.  
 In this way, the $\csc^{2}{(\chi)}$ part of the
trigonometric Rosen-Morse potential becomes part of the kinetic energy
operator on $S^3$, while the $\cot(\chi)$ term acquires meaning of perturbation of
the free quantum rotation on $S^{3}$. The case is quite interesting
indeed, especially for the descriptions of systems of charged particles,
be them of electric, or color-electric charges. The issue is that on
closed manifolds the fields generated by single charges do not 
respect   Gauss's theorem \cite{Landau:1982dva}. Closed manifolds are necessarily charge neutral
and for the fundamental degrees of freedom there are charge-dipoles. From a
technical point of view, this situation generalizes the case of the
ideal gas of di-atomic molecules from two- to three dimensional surfaces of constant curvatures.  Ideal Bose-Einstein and Fermi-Dirac quantum gases in any dimension  have been extensively studied in the literature, see \cite{Valluri} as a Representative example.

Systems
of this kind can appear in the electrodynamics of neutral fluids, for
example, or in Quantum Chromodynamics (QCD), known to describe color-electric
charge neutral systems. Such are for example all the so far observed
mesons, which are composed by equal number of quarks and anti-quarks,
and also all the detected baryons, composed by three quarks in a
color-neutral state. The  non-observability of color-electric
charges in QCD is an important aspect of the fundamental phenomenon known under the name of ``color confinement''. It can
easily be verified that the cotangent function solves the Laplace
equation on $S^{3}$ which establishes it as a harmonic function on
this manifold, quite in parallel to the inverse distance potential in
plane Euclidean space, a reason for which when
\begin{equation}
\label{6}
2b = \alpha Z,\ \ \ \ \ \ \alpha = \frac{e^{2}}{4\pi \hbar c \epsilon_{0}},
\end{equation}
with $\alpha$ being the fine-structure constant, and Z the electric
charge, one refers to the cotangent function as to the ``curved
Coulomb'' potential \cite{Blinder:1996sm}. The remarkable point of the
$ \cot (\chi)$ function is that on $S^3$ it represents a potential generated by a charge-dipole configuration.  
Similarly, the parametrization of ref.
\cite{Kirchbach:2016scz} 
\begin{equation}
\label{7}
2b = \alpha_{s}N_{c},
\end{equation}
where $ \alpha _{s}$ is the (running) strong coupling ``constant'', and
$N_{c}$ the number of colors, appears suited for studies of the color
neutrality of hadrons.

The version of
$V_{tRM}^{\left( l + 1,b \right)}\left( \chi \right)$, to be termed to as ``color-dipole potential'',  is suited
for the description of multiplicities of states in a level which appear  in charge neutral systems
whose orbital angular momenta, $l$ and node numbers, $n$, of the
wave functions, are varying according to $(l\  + \ n)\  = \ k$, with
$k$ non negative integer. Such multiplicities  are observed besides in the
Hydrogen atom, also in a variety of meson spectra \cite{Kirchbach:2016scz,Afonin:2007sv}.

In contrast, the general version,
$V_{\text{tRM}}^{\left( \bar a,b \right)}\left( \chi \right)$ with real $\bar a$
parameter, is suited for spectra in which both charge neutrality and
state multiplicities are absent, as emerging in atomic and molecular systems.
Both versions of the potential under discussion have been studied by the
super-symmetric quantum mechanics and its exact solutions can be found
among others in \cite{Compean:2005cc,Raposo:2007qma}.
Notice that a meson, in being constituted by a quark ($q)$ and an anti-quark $(\bar q)$, is a color dipole.  A gluon, $(g)$, and an anti-gluon ($\bar g)$ represent another type of a color dipole, which also qualifies as a  part of the internal structures of mesons. The potential in (\ref{4}) possesses several essential traits of strong interactions. For example, at short distances it is Coulomb-like, while at larger distances it becomes ``stringy'' as it becomes  dominated by a linear term, all properties well visible by its Taylor series expansion around origin (``North pole''),

\begin{eqnarray*}
-2b\cot \chi &\approx&  -\frac{2b}{\chi} +\frac{4b}{3}\chi .... , \quad 
\chi=\frac{\stackrel{\frown}{r}}{R}.
\end{eqnarray*} 

However, in contrast to the Coulomb- plus linear interaction (known in the literature as ``Cornell potential''), the cotangent potential is of finite range and describes systems confined to finite  hyper-spherical volumes.  
In effect, the potential under consideration provides a reasonable  parametrization of  the internal  structure of  mesons as $(q\bar q)$ color-dipoles moving in the field generated by 
$(g\bar g)$ color dipoles and justifies employment of the equations (\ref{4}) and (\ref{7}) in the description of the thermodynamical properties of mesonic  quantum gases. 

While the trigonometric Rosen-Morse potential in Eq. (1) has been frequently
employed in spectroscopic studies, its thermodynamical properties are
much less known \cite{Blinder:1996sm,Abu-Shady:2019dqv}. The reference \cite{Blinder:1996sm} is  exclusively
devoted to the evaluation of the canonical partition function of the
curved Coulomb potential alone, while in \cite{Abu-Shady:2019dqv} the authors calculate
the partition function of the potential in (\ref{1}) for a fixed real value of
the $\bar a$ parameter, thus letting the summation run only over
the node number, $n$. As explained above, the latter choice does not
allow the $\csc^{2}{(\chi)}$ term to become part of the
Laplace-Beltrami operator on $S^{3}$, a reason for which the aspects
of state multiplicities and charge neutrality cannot be addressed. It is
the goal of the current work to go beyond the aforementioned approaches and

\begin{itemize}
\item improve the expression for the canonical partition function 
reported in \cite{Blinder:1996sm},

\item  work out the canonical and grand canonical
partition functions relevant for color-electric charge neutral systems with state multiplicities in a level,

\item work out various thermodynamic functions related to the aforementioned partition functions, 

\item  study the influence of the finite volume of the space occupied by the quantum gas, fixed by the volume of the hyper-sphere.

\end{itemize}
More precisely,  we shall calculate the partition  functions for non-negative
integer $\bar a$ values, i.e. for $\bar a = l + 1$, with
$l = \ 0,\ 1,\ 2,\ ...,\ k$, extend the summation over
$\left( l\  + \ n \right) = k$, and investigate the thermodynamic
properties of the trigonometric Rosen-Morse potential on these grounds. 
All in all, the system under consideration is akin to one consisting of meson-meson molecules
which behaves like a Bose-Einstein gas, which effectively conserves the total particle number. The standard definition of chemical potential follows, see \cite{Cook:1997cd}. A similar analysis for a similar bosonic system of a pion gas can be found in this work~\cite{Begun:2006gj}.

The text is structured as follows. The next section is devoted to the canonical 
partition function. In section 3 we calculate and plot all the
thermodynamic functions following from this canonical partition function. In
section 4 the grand canonical partition function is elaborated together
with the corresponding pressure function. The results are illustrated on the examples of charmonium and $f_0$ mesons. In this way we
lay down  the mathematical foundations for applications in particle physics of the color-dipole potential  introduced above. The text closes with a brief summary  of the
results.

\section{The canonical  partition function $Z(\beta,b,R)$ of the color-dipole potential}
\label{sec:2}

To evaluate the canonical partition function for the potential in Eqs. (\ref{4})-(\ref{5}),  and (\ref{7}) on $S^{3}$, the linear energy in units of MeV is read off from eq.~(4) 
as,
\begin{equation}
\label{8}
E_k=\frac{\hbar^{2}c^{2}}{2Mc^{2}R^{2}}\ \left( \left( k + 1 \right)^{2} - \frac{b^{2}}{\left( k + 1 \right)^{2}} \right),
\end{equation}
where $Mc^{2}$ is the reduced mass of the two-body system under
consideration. Then the so called  ``rotational temperature'' expresses as, $T_{rot} = \frac{\hbar c}{R}$. The kinetic energy contribution is represented in Eq.~\ref{8} by the $(k+1)^2$ term while the potential
energy is contained in the $b$ dependent term. As we will see next, the kinetic energy is indeed the most
significant contribution to the partition function as the potential energy term decreases
very rapidly beyond the first excited state.

The  expression for the partition function which does not account for 
multiplicity of states in a level,  here denoted by $Z(\beta,b,R)$, is given by. 
\begin{eqnarray}
Z^{}(\beta,b,R)&=&\sum_{k=0}^\infty   e^{-\beta (E_{k}-E_0)},
\label{Z-nodeg}
\end{eqnarray}
where the thermodynamic $\beta$ constant is defined as $\beta = \frac{1}{K_{B}\text{ T}}$, with $K_{B}\ $standing for the Boltzmann constant.
This function can be approximated according to,
\begin{eqnarray}
Z^{}(\beta,b,R) &\approx&  e^{\beta E_0}\int _0^\infty e^{-\beta x^2}
dx\nonumber\\
&=&\frac{
e^{\beta E_0}\sqrt{\pi}~\mbox{erfc}(
\sqrt{\gamma\beta})}{2\sqrt{\gamma\beta}},\ \ \  Re(\sqrt{\gamma\beta})>0, \qquad
\label{Z_nondeg} 
\end{eqnarray}
where
\begin{eqnarray} 
&\quad & \gamma=\frac{\hbar^2c^2}{2Mc^2 R^2}, 
\label{Z_nondeg_1} 
\end{eqnarray}
and  $\mbox{erfc}\left( u \right)$ stands for the complementary error function \cite{Arfken:2013mph}.

\begin{figure}
\center	
\includegraphics[width=0.75\textwidth]{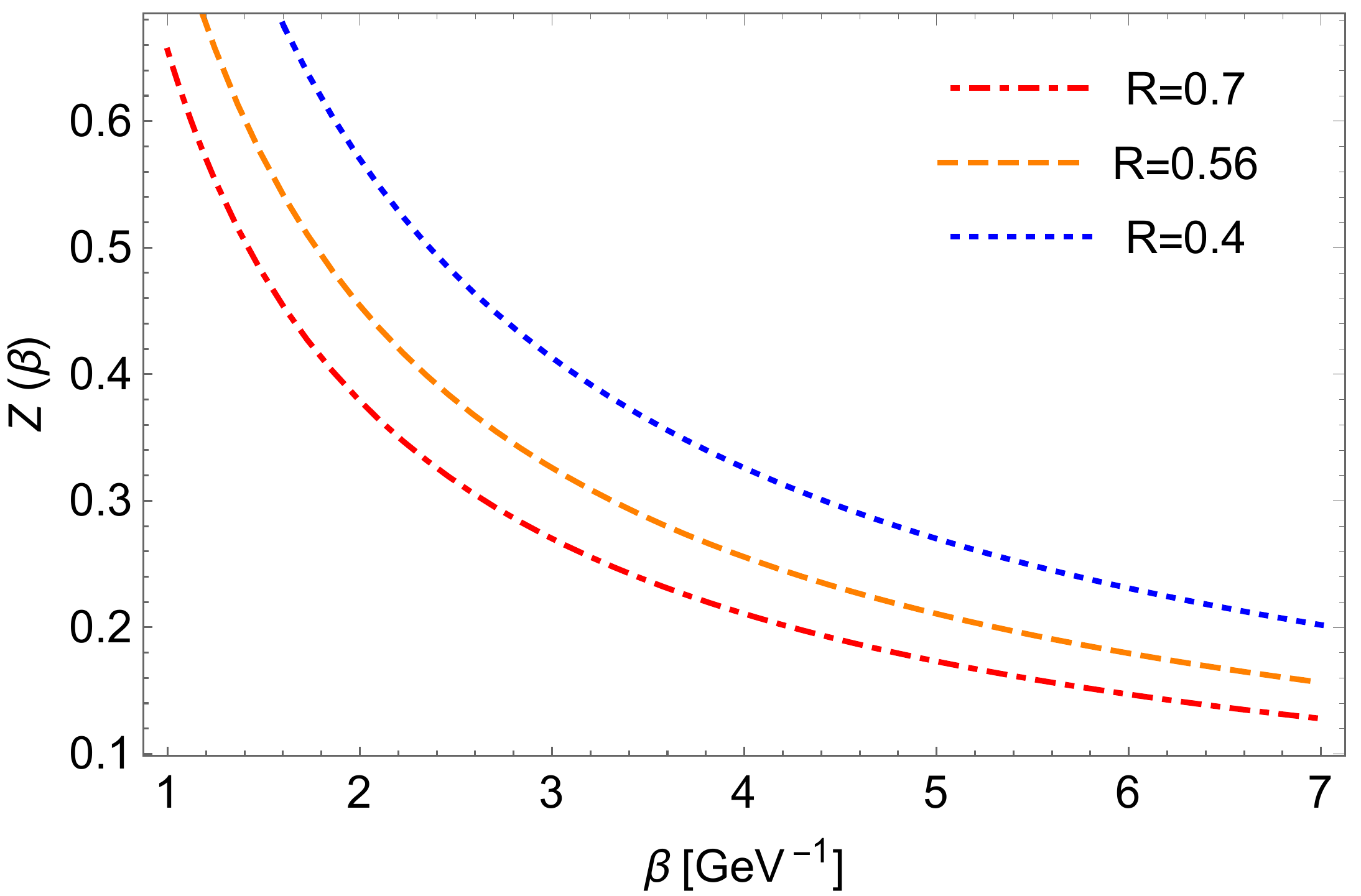}
\caption{\label{F-1-Non-degenerate-PF} The canonical partition function in Eq. (\ref{Z_nondeg}) as a function of $\beta$, with $\beta$ varying from 1 GeV$^{-1}$ to 5 GeV$^{-1}$.
The parameters $R$ and $\alpha_s$ take the values $0.7, 0.56, 0.4$ fm associated to $0.66\pi, 0.2, 0.1$, respectively. $f_0(500)$ and charmonium values are denoted in red and orange colors.}
\end{figure}

\begin{figure}[htpb!]
\center	
\includegraphics[width=0.75\textwidth]{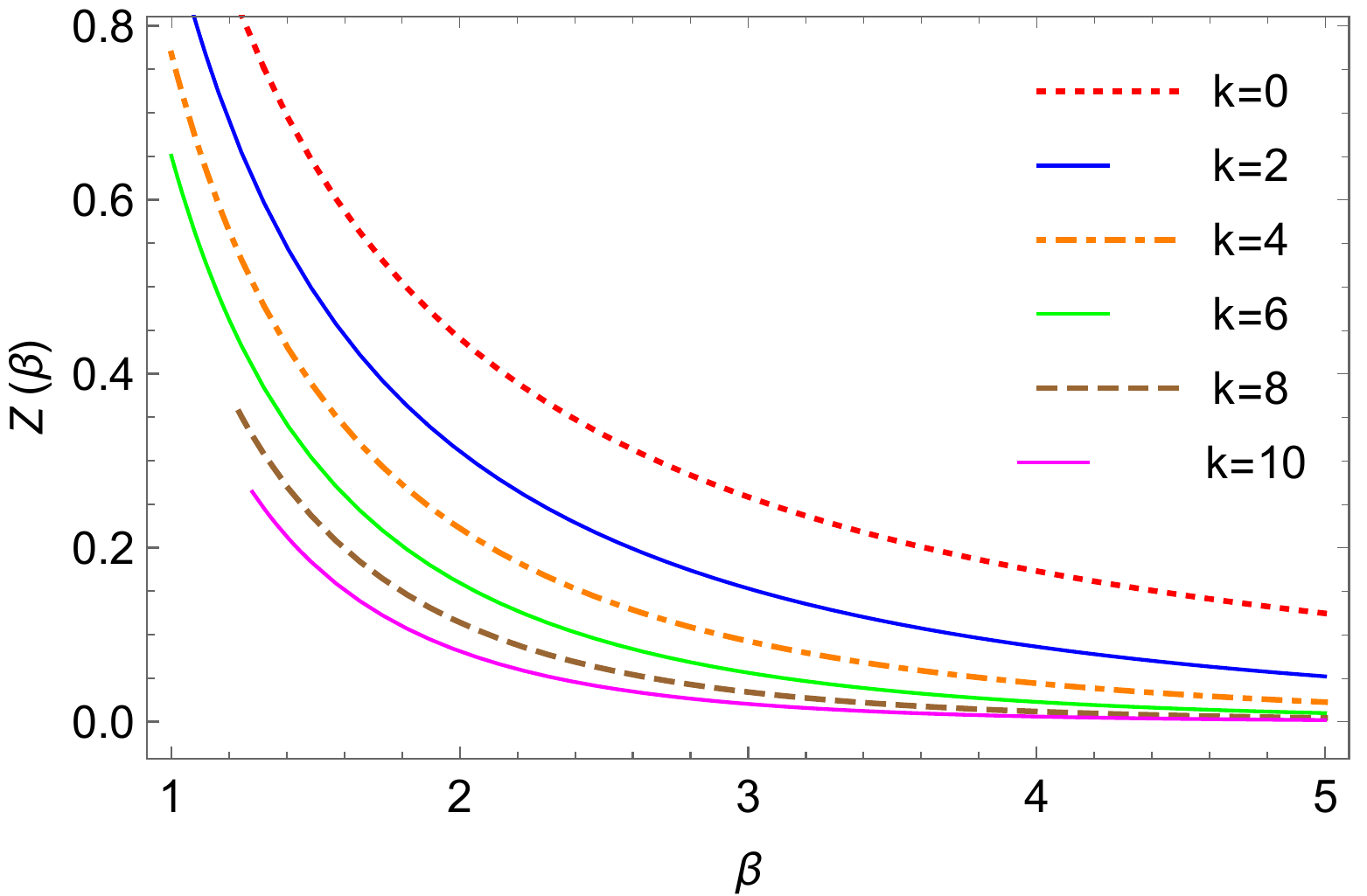}
\caption{\label{Z-contributions} Partial contributions to the
partition function from the $k$-term in Eq.~(\ref{9}). It can be seen that already for $k=10$ the contribution from this
term is negligible, justifying the continuous approximation.}
\end{figure}

In Fig. \ref{F-1-Non-degenerate-PF} we display the canonical partition function that ignores 
the multiplicities of the states in the levels.
The definition of  the  partition function that accounts for states multiplicities, again denoted by $Z(\beta,b,R)$ for the sake of not overloading the presentation with  too much different notations, and concerning the potential under consideration, is standard  \cite{Hoover:2012csm} and given by,
\begin{equation}
\label{9}
Z\left( \beta,b,\ R \right) = \sum_{k = 0}^{\infty} g_{k} e^{- \beta(E_{k} - E_{0})},\quad k=\ell +n,\quad g_k=(k+1)^2,
 \end{equation}
where the energy multiplicity  in the levels is $\left( k + 1 \right)^{2}$-fold. Now, Eq. (\ref{9}) can be evaluated as,
\begin{eqnarray}
\label{10}
Z\left( \beta,b,\ R \right) &=& e^{\beta E_0}\sum_{k = 0}^{\infty}\left( k + 1 \right)^{2} e^{- \beta E_{k}} \nonumber\\ 	
				     &\approx& 
e^{\beta E_{0}} \int_{0}^{\infty} \left( x + 1 \right)^{2} e^{- \beta E(x)}d x,
\end{eqnarray}
where the discrete variable $k$ has been approximated by a continuous one, denoted by $x$, the infinite sum has been approximated by integration, and the discrete energy $E_k$ has become the function, $E(x)$. 
Figure~\ref{Z-contributions} shows that due to the exponential factor in the expression for the partition function of Eq.~(\ref{9}) rapidly supresses higher order contributions to the sum,  with only a few terms needed to be very close to the above continuous approximation. Along this line, in Ref.\cite{Blinder:1996sm} , the partition function has been worked out for the case of $b = \alpha Z/2$ and corresponds to the Hydrogen atom
on $ S^{3}$.  With that, the partition function presents itself as,
\begin{equation}
Z\left( \beta,b,\ R \right) \approx  e^{\beta E_{0}} \int_{0}^{\infty} (x+1)^{2} 
e^{- a (x+1)^{2} +\frac{p}{(x+1)^{2}}} dx,
\label{Gl14} 
\end{equation}
where,  we introduced the new notations of 
\begin{equation}
a = \beta\ \gamma, \qquad p = \beta\  \gamma\  b^2.
\label{prmtrs_erfc}
\end{equation}
Now the integral can be evaluated by integration by parts yielding,
\begin{eqnarray}
\label{12}
 \int_{0}^{\infty} (x+1)^{2} e^{- a(x+1)^{2}+\frac{p}{(x+1)^{2}}} dx &&\nonumber\\
=\int_{0}^{\infty}{(x+1)^{2}\ e^{- a(x+1)^{2}}} e^{\frac{p}{(x+1)^{2}}}dx &&\nonumber \\ 
				     = \frac{- 1}{2a}\int_{0}^{\infty}(x+1)e^{\frac{p}{(x+1)^{2}}} d(e^{- a(x+1)^{2}}) &&\nonumber \\ 	
				     =\frac{- 1}{2a}(x+1)e^{\frac{p}{(x+1)^{2}}} e^{- a(x+1)^{2}}|_{0}^{\infty}&& \nonumber \\ 
				     + \frac{1}{2a}\int_{0}^{\infty}e^{- a(x+1)^{2}} d((x+1)e^{\frac{p}{(x+1)^{2}}})&& \nonumber \\ 	
				      = 
\frac{1}{2a}e^{p-a}  + \frac{1}{2a}\int_0^\infty e^{-a(x+1)^2}e^{\frac{p}{(x+1)^2}}d(x+1) && \nonumber\\
+\frac{2p}{2a}
\int_0^\infty  e^{- a(x+1)^{2}+\frac{p}{(x+1)^{2}}} d\left(\frac{1}{x+1}\right).&&\nonumber \\  
\end{eqnarray}
The argument of the exponential term under the integration sign can be represented as,
\begin{eqnarray}
\label{13}
- a(x+1)^{2} + \frac{p}{(x+1)^{2}} &=& -  z^{2} + 2i\sqrt{ap},  \qquad z = \sqrt{a}(x+1)  +  i \frac{\sqrt{p}}{(x+1)},
 \end{eqnarray}
where $z$ is a complex number.
Hence, substituting Eq. (\ref{12}) and Eq. (\ref{13}) into Eq. (\ref{Gl14}), the partition
function becomes,
\begin{eqnarray}
\label{14}
Z\left( \beta,b,\ R \right) &\approx& \frac{1}{2a} +\frac{1}{2a} e^{\beta E_{0}} \int_{0}^{\infty} e^{-z^{2} + 2i\sqrt{ap}}d(x+1) \nonumber \\
				  & &  + \frac{2p}{2a}e^{\beta E_{0}} \int_{0}^{\infty} e^{-z^{2} + 2i\sqrt{ap}}d\left(\frac{1}{x+1}\right) \nonumber \\
				     &\approx&
\frac{1}{2a}  +  \frac{1}{2a} e^{\beta E_{0}} \int_{0}^{\infty} e^{-z^{2} + 2i\sqrt{ap}}d(x+1) \nonumber \\
& & + \frac{1}{2a}e^{\beta E_{0}} \int_{0}^{\infty} e^{-z^{2} + 2i\sqrt{ap}}d\left(\frac{2p}{x+1}\right), \nonumber
\end{eqnarray}
equivalently,
\begin{eqnarray}
Z(\beta,b,R) &\approx& \frac{1}{2a} 
+ \frac{1}{2a}e^{\beta E_{0}}e^{2i\sqrt{a p}}\int_{0}^{\infty} e^{{-z}^{2}} d\left(x+1 + \frac{2p}{x+1}\right). \nonumber \\
 \end{eqnarray}
 Then, with the aid of Eq. (\ref{13}), we conclude on
\begin{eqnarray}
\left(x+1 + \frac{2p}{x+1}\right) &=& \frac{1}{2}\left(\frac{1}{\sqrt{a}} - 2i\sqrt{p}\right)\ z + \frac{1}{2}\left(\frac{1}{\sqrt{a}} + \ 2i\sqrt{p}\right) z^{*}. 
\label{15}
\end{eqnarray}
In effect, the partition function can be expressed in terms of the
erfc($u$) function of complex argument, $u\in {\mathcal C} _2$. In so doing, Eq. (\ref{14}) reduces to,
\begin{eqnarray}
\label{16}
Z\left( \beta,b,\ R \right) &\approx&  \frac{1}{2a} + \frac{1}{4a} e^{\beta E_{0}}  \nonumber\\
&\times & \big[(\frac{1}{\sqrt{a}} - 2i\sqrt{p})e^{2i\sqrt{ap}} \int_{\Gamma}e^{-z^{2}}dz \nonumber \\
&&+(\frac{1} {\sqrt{a}} + 2i\sqrt{p})e^{-2i\sqrt{ap}} \int_{\Gamma} e^{- (z^{*})^{2} }dz^{*}\  \big],\nonumber\\
\end{eqnarray}
where $\Gamma$ is some path on the complex plane starting from $ u=0 $ and ending in $u \rightarrow \infty$. Thus, the partition
function in Eq. (\ref{10}) takes the following shape,
\begin{eqnarray}
\label{17}
Z\left( \beta,b,\ R \right) \approx \frac{1}{2a}  + \frac{1}{4a}e^{\beta E_{0} }\frac{1}{8 a^{3/2}} \times \nonumber \\
 \big(  \sqrt{\pi} e^{- 2\sqrt{a}\sqrt{- p}}  \big( - \text{erfc}\big( \frac{\sqrt{- p}}{x + 1}  - \sqrt{a}\big( x + 1 \big) \big) \nonumber \\
 + e^{4\sqrt{a}\sqrt{- p}}\big( \text{erfc}\big( \sqrt{a}\big( x + 1 \big) + \frac{\sqrt{- p}}{x + 1} - 1 \big)\  + 1 \big) \nonumber \\
 - 2\sqrt{\pi}\sqrt{a}\sqrt{- p}\text{\ e}^{- 2\sqrt{a}\sqrt{- p^{2}}}\big( \text{erfc}\big( \frac{\sqrt{- p}}{x + 1} - \sqrt{a}\big( x + 1 \big) \big) \nonumber \\
 e^{4\sqrt{a}\sqrt{- p}}\big( \text{erfc}\big( \sqrt{a}\big( x + 1 \big) + \frac{\sqrt{- p}}{x + 1} \big) - 1 \big) + \ 1 \big) \nonumber  \\
 - 4\sqrt{a}\big( x + 1 \big)\text{\ e}^{- a\big( x + 1 \big)^{2} + \frac{p}{\big( x + 1 \big)^{2}}} \big)  + c ...\big). \nonumber  \\
 \end{eqnarray}
Our expression for the canonical partition function in (\ref{17}) differs from the one reported in \cite{Blinder:1996sm} through the first term which in \cite{Blinder:1996sm} takes the value of one, because there the integral in 
Eq.~(\ref{12}) has been evaluated  for $(x+1)\in [0,\infty]$, while we evaluate it for $x\in [0,\infty]$, as it should be because the $k$ parameter in (\ref{10}) can not become negative. The additive
$1/(2a)=1/(\beta b\gamma) $ correction to the canonical partition functions is important insofar as  without it the partition function does  not come out as a monotonous function of $\beta$.

Furthermore, as a further attempt to simplify the integration of Eq. (\ref{10}), an approximate  treatment can be suggested by using the Taylor series
expansion \cite{Wen:2019s}. Along this line, first the integration can be expanded around
$x= 0$, according to,
\begin{eqnarray}
\label{18}
Z\left( \beta,b,\ R \right) &=& \frac{1}{2a}+ e^{\beta E_{0}\ }\big( e^{p - a} -2k\big( e^{p - a}\big( a + p + 1 \big) \big)  \nonumber \\
&+&k^{2}\ e^{p - a}\ \big( 2a^{2} + a\big( 4p - 5 \big)\nonumber\\
 &+& 2p^{2} - \ p + 1 \big) -\frac{2}{3}k^{3}{(e}^{p - a}\ \big( 2a^{3} + a^{2}\big( 6p - 9 \big)\nonumber\\
& +& 6a(2p^{2} - p + 1 \big) + p^{2}(2p + 3))\ ) \nonumber \\
 &+&\frac{1}{4}k^{4}e^{p - a}\big( {4a}^{4} + {4a}^{3}\big( 4p - 7 \big)\nonumber\\
& +& {3a}^{2}\big( {8p}^{2} - 12p + 13 \big) \nonumber  \\
& +&2a\big( 8p^{3} + 6p^{2} + 3p - 3 \big) + p^{2}\big( 4p^{2} + 20p + 15 \big) \big) + O(x^{5}) \big). \nonumber \\
 \end{eqnarray}
Back to the partition function in  Eq.~(\ref{Gl14}), we notice that with the
increase of $x$, the contribution to the argument of the exponential  provided by the term defined by the  $p$ parameter  can
be ignored compared to the term proportional to ${(x + 1)}^{2}$. In addition, as visible from their definitions in (\ref{prmtrs_erfc}) in combination with (\ref{7}), 
also the $p$ parameter is significantly smaller than the $a$ parameter, especially in the QCD regime of the asymptotic freedom where $\alpha_{s} \rightarrow 0$. For this case, the canonical partition
function describes an ideal gas of color-electric dipoles, though the interaction has left its footprint through the presence of the  $p$ parameter in the $e^{\beta E_0}$ factor with $E_0=(a-p)$. Correspondingly,
the canonical partition function under investigation becomes real and can be is approximated as,
\begin{eqnarray}
\label{Z_deg}
Z\left( \beta,b,\ R \right)  &\cong& \frac{1}{2a} + 4\ e^{- 3\beta E_{0}} + \int_{0}^{\infty}\left( x + 1 \right)^{2}e^{- \beta\left[\left( x + 1 \right)^{2}E_{0} - E_{0}\right]}dx    \nonumber \\
&\cong& \frac{1}{2a} + 4\ e^{- 3\beta E_{0}} + e^{\beta E_{0}}\int_{0}^{\infty}{\left( x + 1 \right)^{2}e^{- \beta E_{0}\left( x + 1 \right)^{2}}}dx  \nonumber \\
&\cong& \frac{1}{2a}+4 e^{- 3\beta E_{0}} + \frac{\sqrt{\pi}}{4}\frac{1}{\left( E_{0}\beta \right)^{\frac{3}{2}}}\ e^{\beta E_{0}}. \nonumber \\
\end{eqnarray}
Hereafter, if not said otherwise, we will consider the thermodynamic functions at fixed $R$ value, and employ  Eq. (\ref{Z_deg}). 
In what follows we suppress the $R$ and $b$ arguments in $Z(\beta,b,R)$  for the sake of simplifying notations, and focus on  the dependence of the 
partition function on $\beta$ alone, just writing $Z(\beta)$. 

\begin{figure}	
\center
\includegraphics[width=0.75\textwidth]{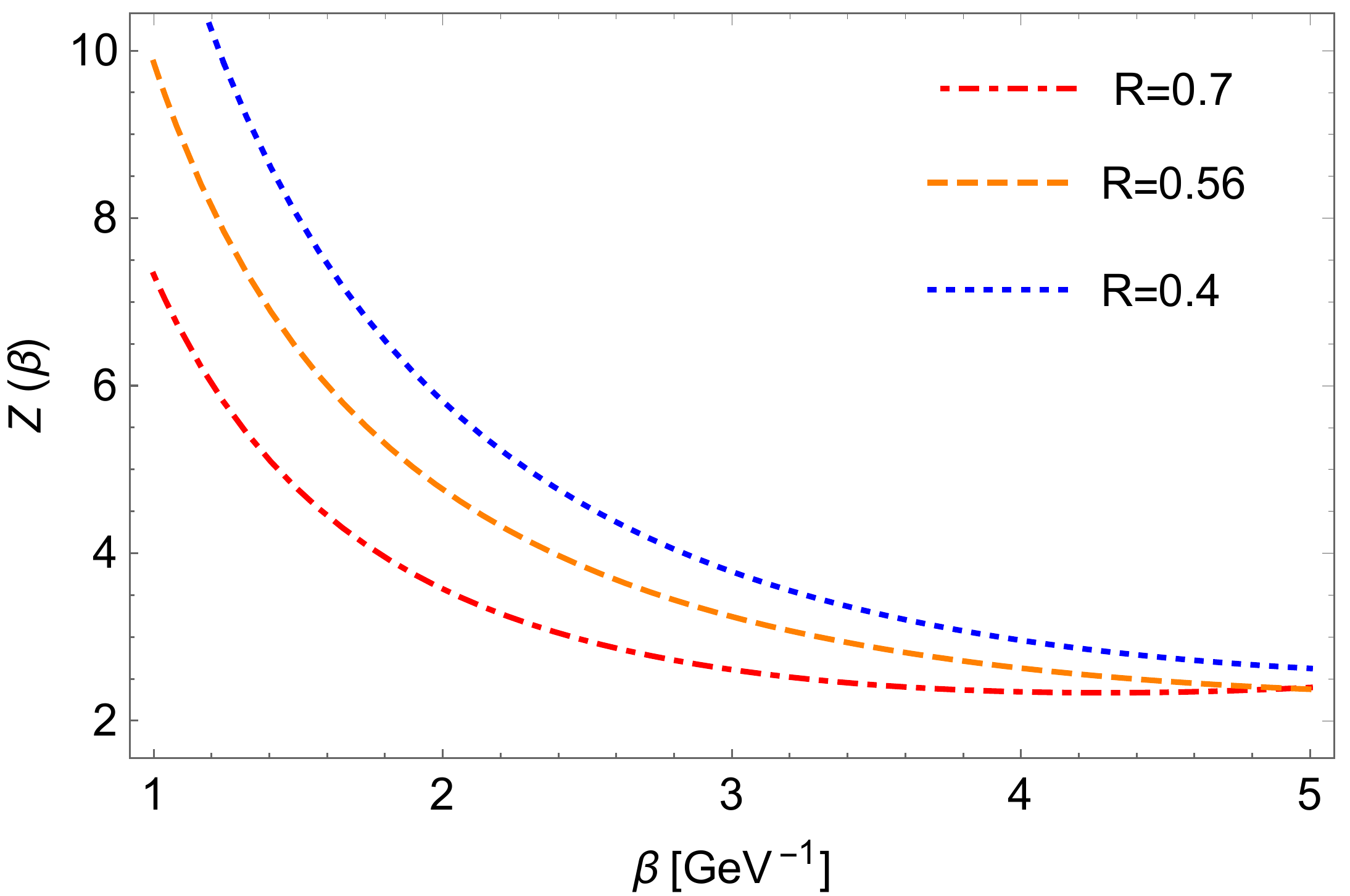}
\caption{\label{F-2-Degenerate-PF} The partition function which accounts for multiplicities in Eq. (\ref{Z_deg}) as a function of $\beta$, with $\beta$ varying from 15 GeV$^{-1}$ to 5 GeV$^{-1}$. The parameters $R$ and $\alpha_s$ take the values $0.7, 0.56, 0.4$ fm associated to $0.66\pi, 0.2, 0.1$, respectively. $f_0(500)$ and charmonium values are denoted in red and orange colors.}
\end{figure}

Now, the moment of inertia, $I$,  is related to rotational temperature as,
\begin{equation}
\label{21}
\frac{\hbar^{2}c^{2}}{2I} = K_{B}T_{\textit{rot}}, \qquad T_{\mbox{rot}}=\frac{\hbar c}{R}.
\end{equation}
with
\begin{equation}
\label{22}
I = Mc^{2}R^{2}.
\end{equation}
In addition, the classical energy $E$ is related to the rotational temperature as,
\begin{equation}
\label{23}
E = K_{B} T_{\textit{rot}}.
\end{equation}
The Figure \ref{F-2-Degenerate-PF}  represents  the canonical partition function in (\ref{Z_deg}) which accounts for the multiplicities. Notice the monotonous fall off of the function with $\beta$, a behavior ensured by the correct additive $1/(2a)$ contribution to (\ref{Z_deg}). In contrast, in \cite{Blinder:1996sm}, one encounters in place of $1/(2a)$ the $1$ constant, instead, in which case the partition functions is not monotonous. 
Comparison of Figs.~\ref{F-1-Non-degenerate-PF} and \ref{F-2-Degenerate-PF} shows that as expected,  the partition function accounting for the state multiplicities  takes significantly larger values relative to the one which ignores them. Similar is the effect of the decreasing volume but it is  notably smaller in size.

Both figures show that the partition functions corresponding to different volumes converge with $R\to \infty$, a tendency  better pronounced by  Fig.~2. 
The  partition functions  shown in  Figs. \ref{F-1-Non-degenerate-PF}, and 
\ref{F-2-Degenerate-PF} are both smooth functions of $\beta$ and compare in shape with 
a similar result reported in Ref. \cite{Abu-Shady:2019dqv}, where 
the authors employed the trigonometric Rosen-Morse potential in the flat space Schr\"odinger equation.
Similar is the behavior of the partition functions of di-atomic molecules  \cite{Ebert:2002pp}, and of  the harmonic-oscillator plus inverse-squared distance potential \cite{Das:2016t} as well of P\"osch-Teller-type potentials~\cite{Yah:2016Oye}. 
 Therefore,  our choice of integer values for the ${\bar a}$ parameter  in the potential in Eq.~(\ref{1}), which was required by the $S^3$ geometry that ensured the desired charge neutrality, does not alter the general behavior of the partition function with the change of $\beta$.

\section{Thermodynamic properties following from the canonical partition function}
\label{sec:3}
This section is devoted to the elaboration of the thermodynamic functions
following from the canonical partition function worked out in the
previous section. According to \cite{Hoover:2012csm}, the main thermodynamic functions such as the internal
energy ($U(\beta)$), the  energy fluctuation
$\left\langle {(\mathrm{\Delta}U(\beta))\ }^{2} \right\rangle$, and the
heat capacity ($C(\beta)$) can be found from the canonical
partition function $Z\left( \beta \right)$ in Eq. (\ref{Z_deg}) as,
\begin{equation}
\label{24}
\left\langle U(\beta) \right\rangle =  \frac{- \partial \ln \left( Z \left( \beta \right) \right)}{\partial \beta}=\frac{A}{B},
\end{equation}
amounting to,
\begin{eqnarray}
 A&=&12E_{0}\ e^{- 3\beta E_{0}} - \frac{\sqrt{\pi}}{4}\ \left( \frac{E_{0} e^{\beta E_{0}}}{\left( \beta E_{0} \right)^{\frac{3}{2}}} \right)
\nonumber\\
&+&\frac{3\sqrt{\pi}}{8}\ \left( \frac{E_{0}  e^{\beta E_{0}}}{\left( \beta E_{0} \right)^{\frac{5}{2}}} \right)
+ \left( \frac{1}{2\gamma \beta^2\ }  \right),
\nonumber\\
B&=&\frac{1}{2\gamma\beta} + 4\ e^{- 3\beta E_{0}} + \frac{\sqrt{\pi}}{4}\frac{1}{\left( E_{0}\beta \right)^{\frac{3}{2}}}\ e^{\beta E_{0}}\ .
\label{25}
\end{eqnarray}
\begin{figure*}[!bpht]
\begin{center}$
\begin{array}{cc}
\includegraphics[width=0.5\textwidth]{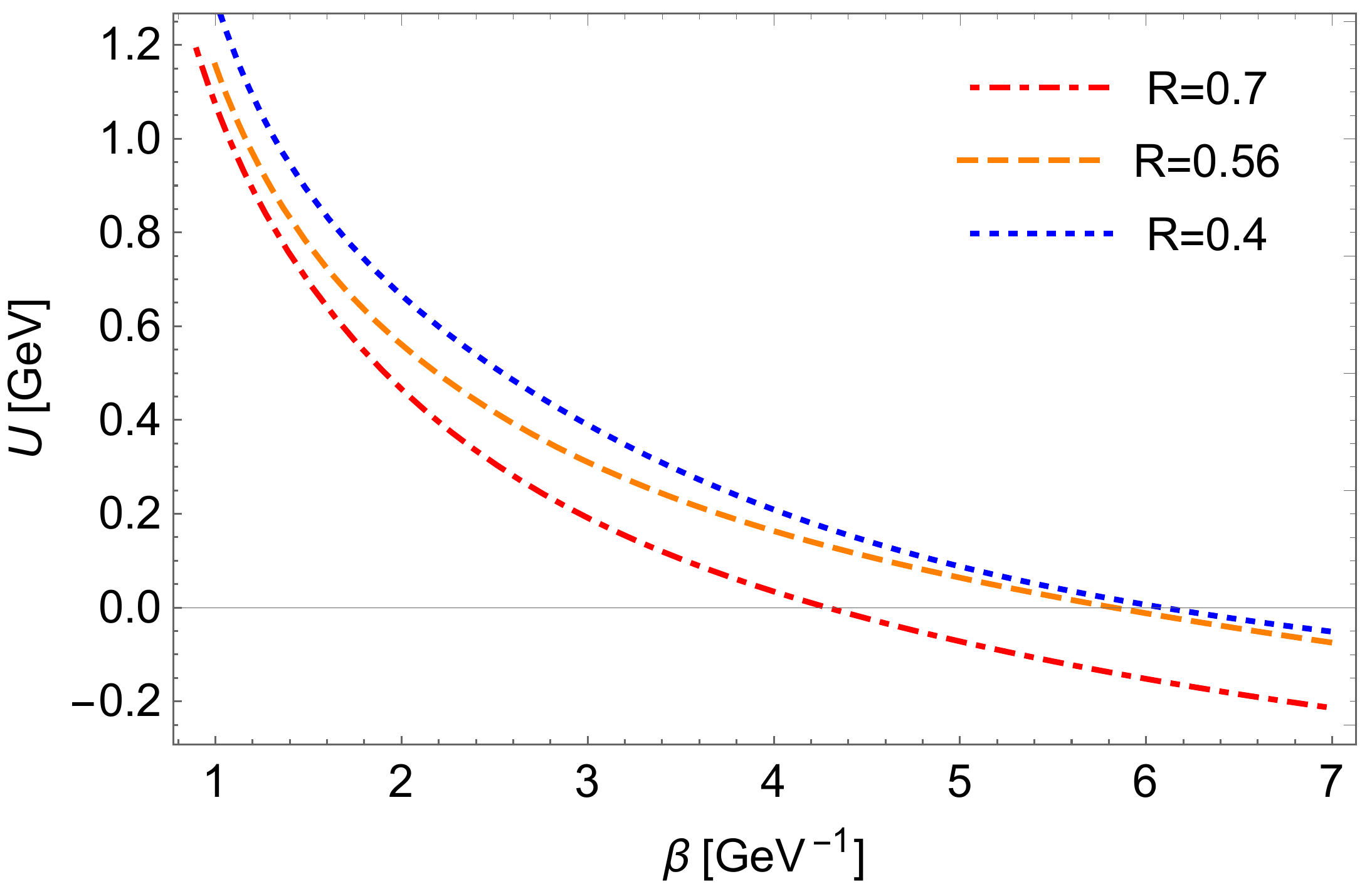} & \includegraphics[width=0.5\textwidth]{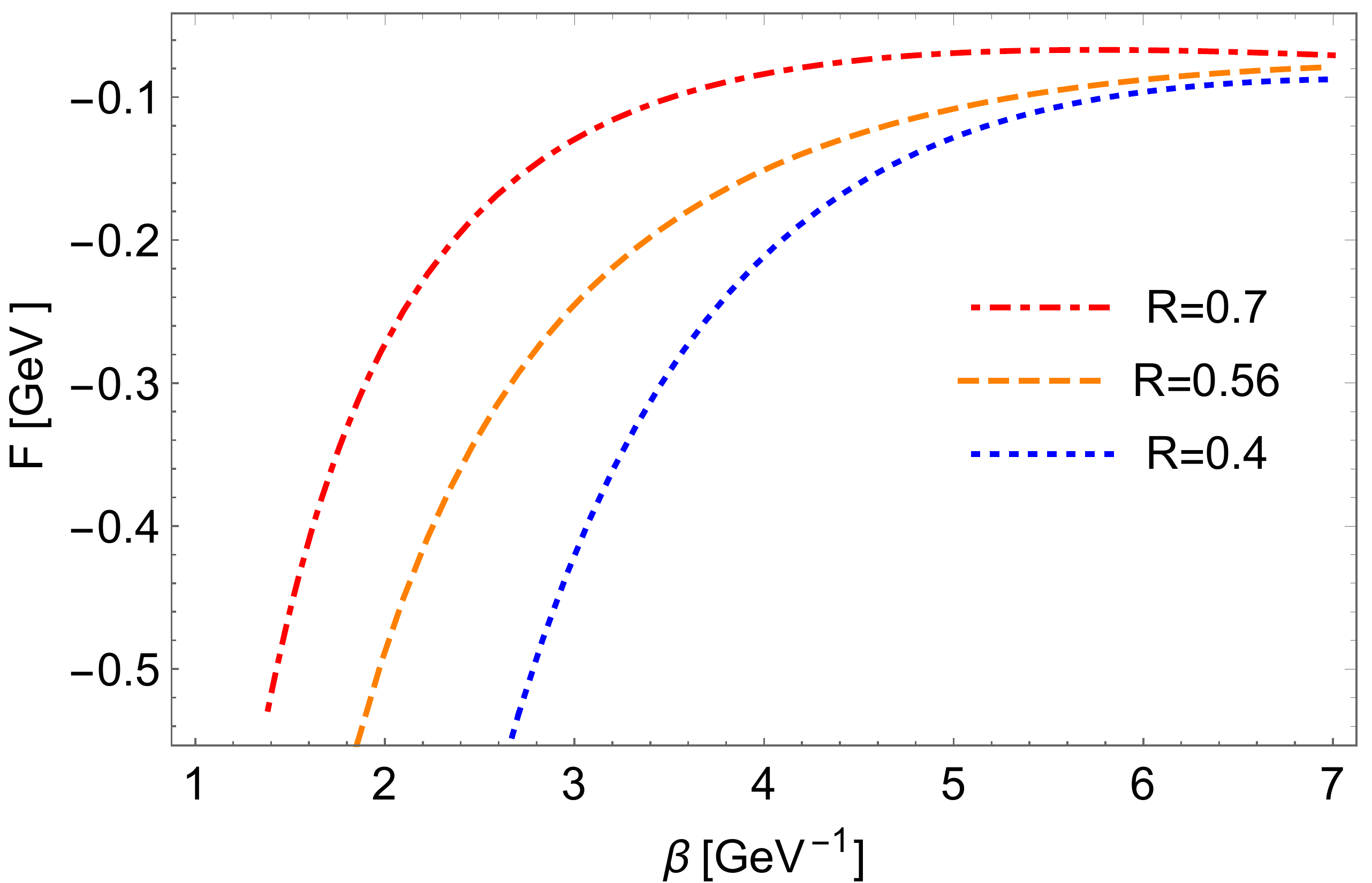} \\
\includegraphics[width=0.5\textwidth]{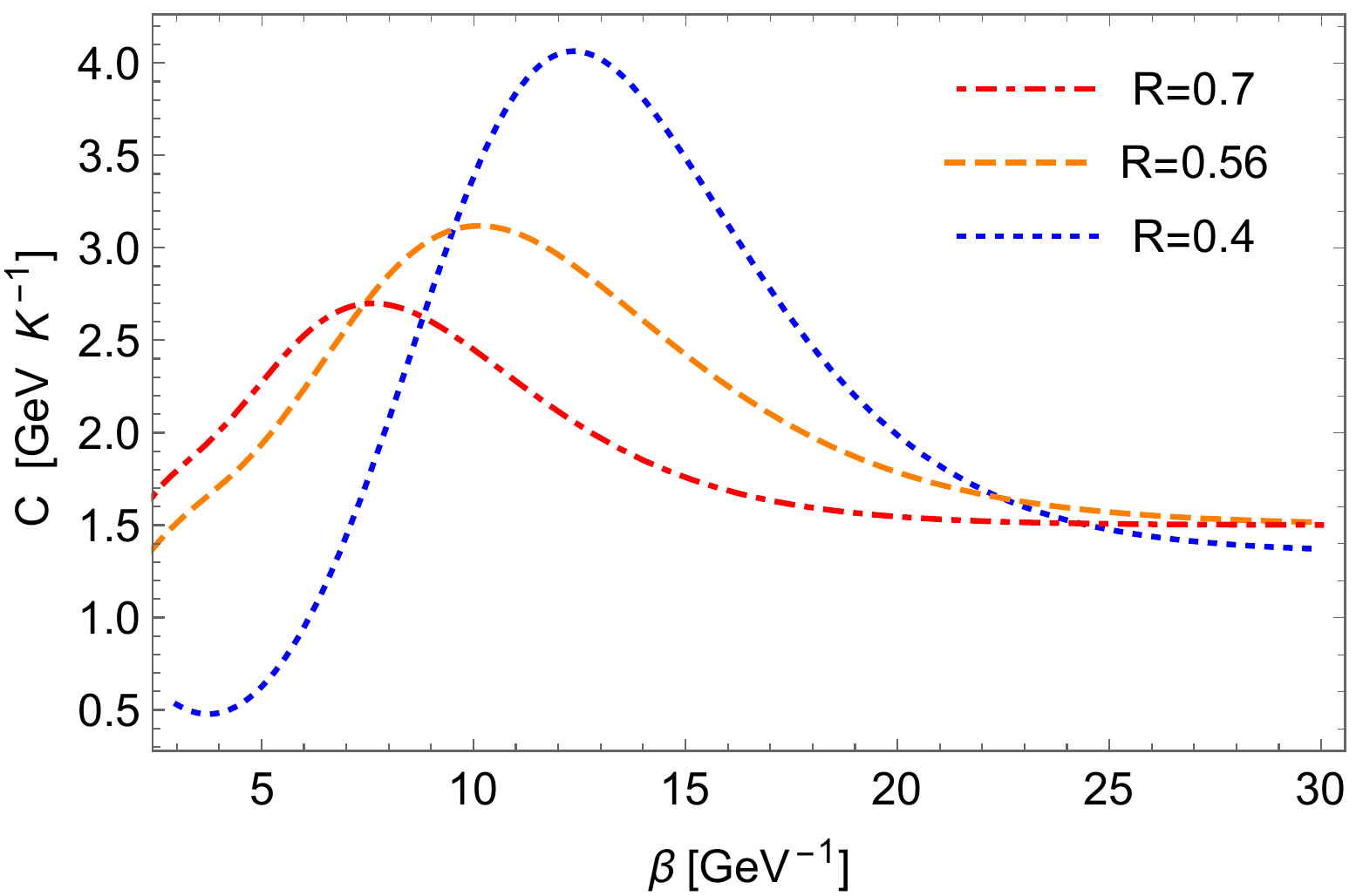} & \hspace{0.45cm} \includegraphics[width=0.4825\textwidth]{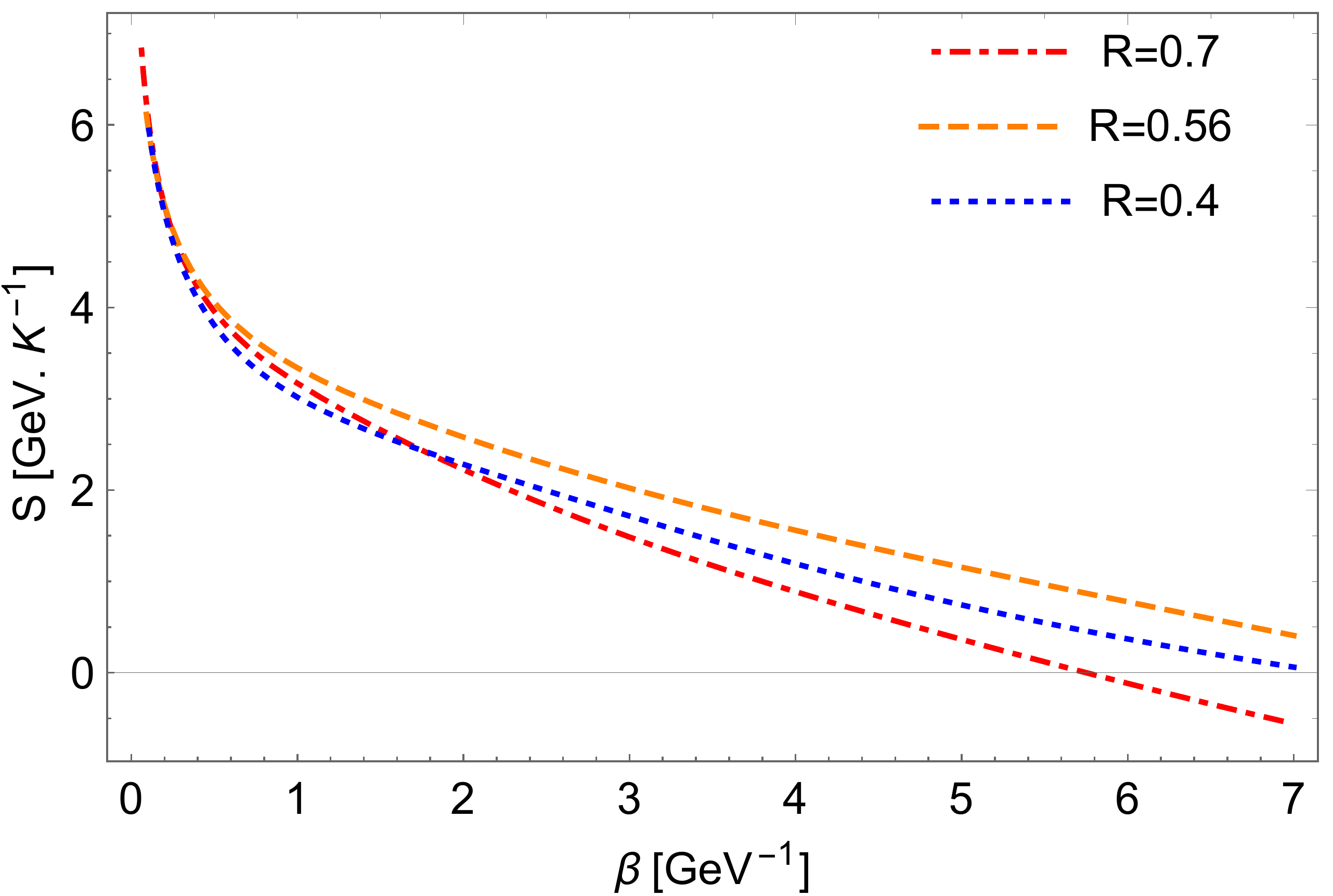}
\end{array}$
\end{center} 
\caption{Thermodynamic quantities derived from $Z\left( \beta,b,\ R \right)$. Top: Internal energy $U$ as derived in Eq. (\ref{24}) and Helmholtz free energy $F$ as derived in Eq.(\ref{29}).
Bottom: Heat capacity $C$ as derived in Eq. (\ref{28}) and entropy $S$ as derived in Eq. (\ref{31}). The parameters $R$ and $\alpha_s$ take the values $0.7, 0.56, 0.4$ fm associated to $0.66\pi, 0.2, 0.1$, respectively. $f_0(500)$ and charmonium values are denoted in red and orange colors.}
\label{UFCS}
\end{figure*}
\begin{figure}[!bpht] 
\center
\resizebox{0.75\textwidth}{!}{%
  \includegraphics{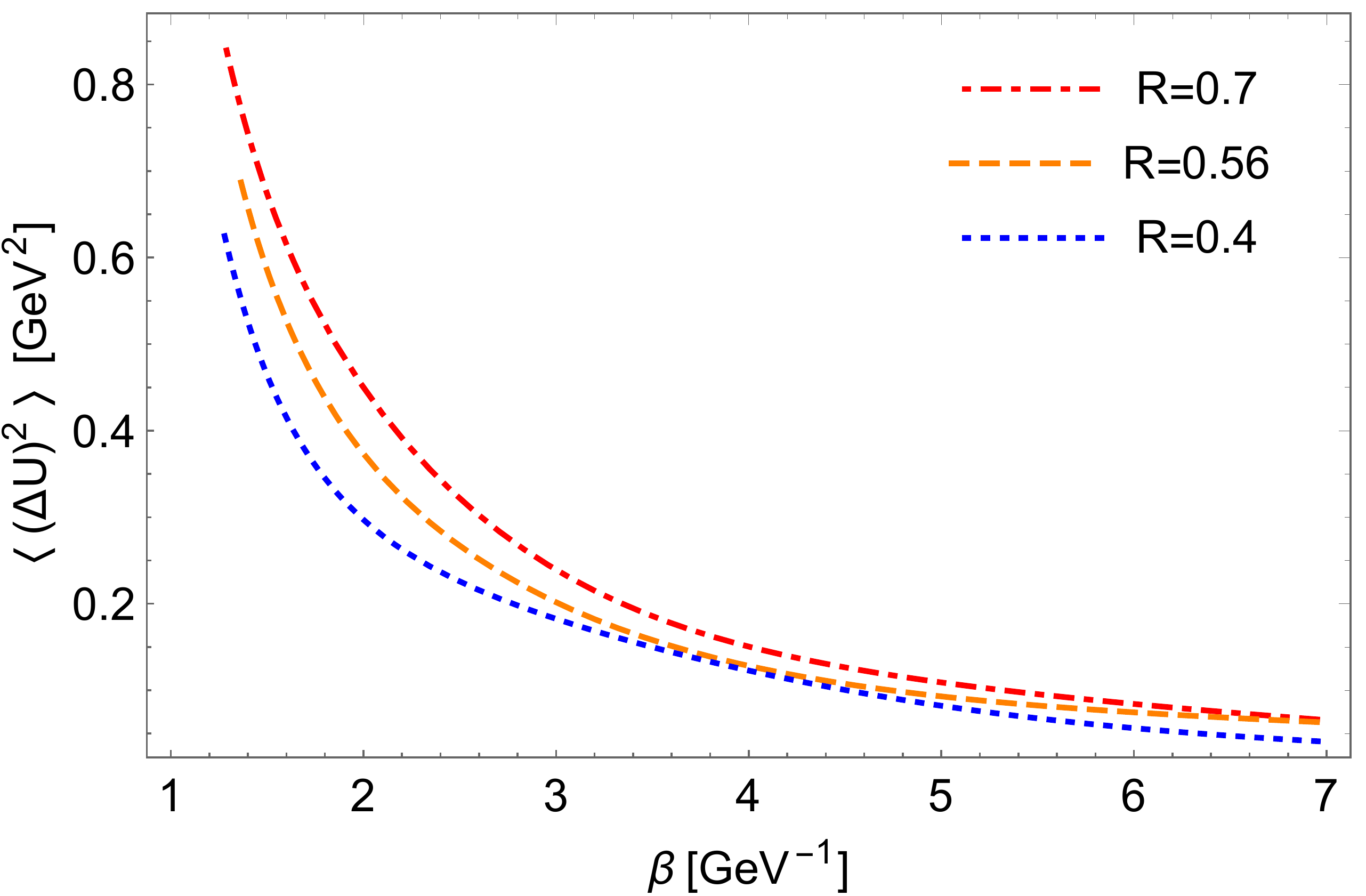}
  }
  \caption{The variation of energy $\left\langle {(U(\beta))\ }^{2} \right\rangle\ $ as derived in Eq. (\ref{26}) as a function of $\beta$. The parameters $R$ and $\alpha_s$ take the values $0.7, 0.56, 0.4$ fm associated to $0.66\pi, 0.2, 0.1$, respectively. $f_0(500)$ and charmonium values are denoted in red and orange colors.}
  \label{DeltaE}
\end{figure}
\begin{figure}[htpb!]	
\center
\includegraphics[width=0.75\textwidth]{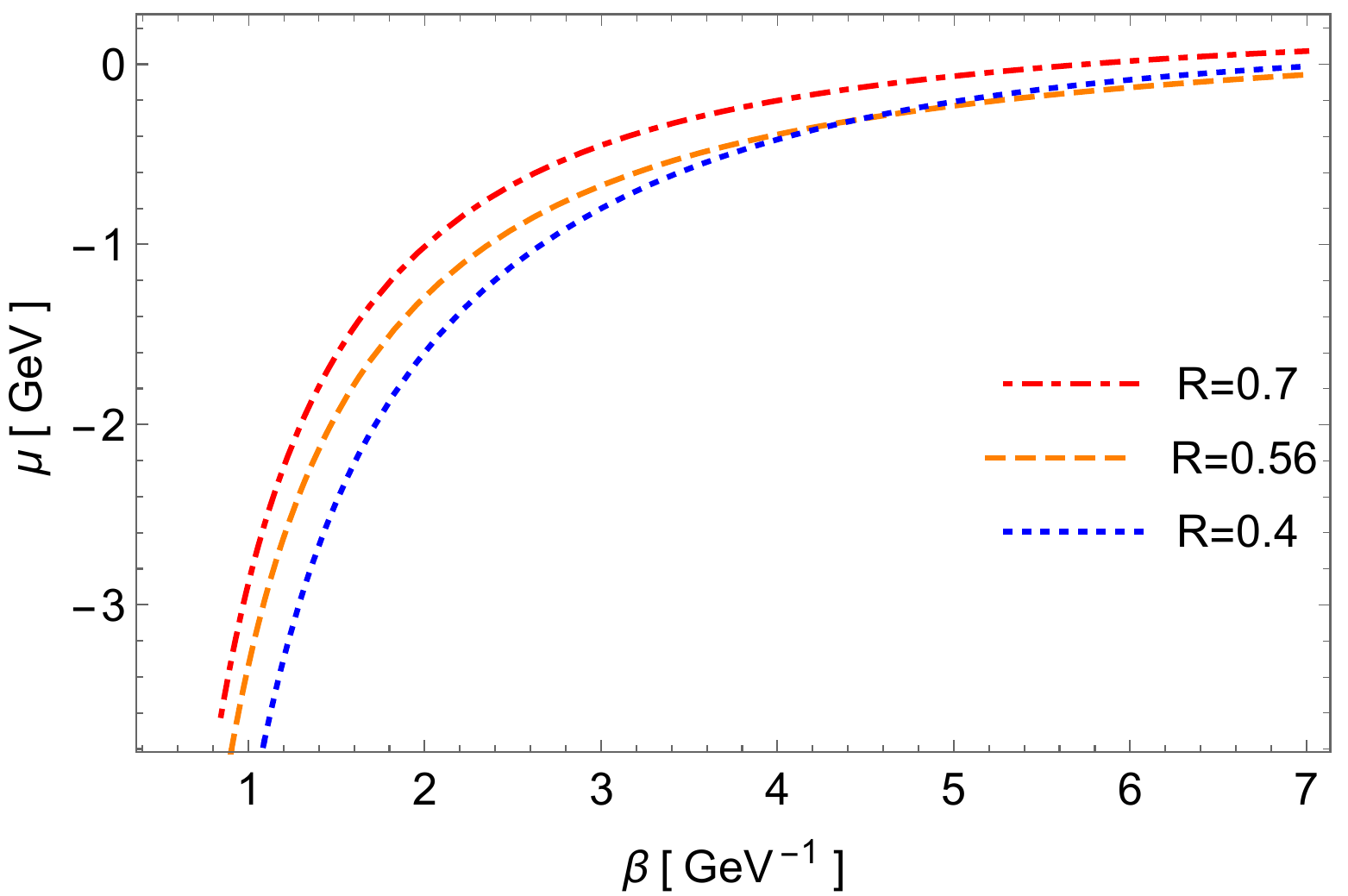}
\caption{\label{a}The chemical potential as derived in Eq. (\ref{35}) as a function of ${\beta}$. $f_0(500)$ and charmonium values are denoted in red and orange colors.}
\end{figure}
Furthermore, the variation of the energy (or "energy fluctuation") is calculated as,
\begin{eqnarray}
\label{26}
\left\langle {(\Delta U(\beta)) }^{2}\right\rangle & =& 
\frac{- \partial\left\langle U(\beta) \right\rangle}{\partial\beta} 
 = \frac{\partial^{2}\ln\left({Z(}\beta) \right)}{\partial\beta^{2}}\\
 &=& -\left( \frac{A}{B}\right)^2+\frac{D}{B}\nonumber,
\end{eqnarray}
with
\begin{eqnarray}
\label{27} 
D=
36 {E_{0}}^{2} e^{- 3\beta E_{0}} +\frac{15 \sqrt{\pi}}{16} \frac{E_{0}^2  e^{\beta E_{0}}}{\left( \beta E_{0} \right)^{\frac{7}{2}}} - \frac{3\sqrt{\pi}}{4} \frac{{E_{0}}^{2} e^{\beta E_{0}}}{\left( \beta E_{0} \right)^{\frac{5}{2}}}+ \frac{\sqrt{\pi}}{4}\frac{{E_{0}}^{2} e^{\beta E_{0}}}
{\left( \beta E_{0} \right)^{\frac{3}{2}}} 
+\frac{1}{\gamma\beta^3}. \nonumber
\end{eqnarray}
Finally, the heat capacity is obtained from Eq. (\ref{24}) as:
\begin{eqnarray}
\label{28}
C(\beta)& =& \frac{\partial U(\beta)}{\partial T} = - {K_{B}\ \beta}^{2}\frac{\partial U(\beta)}{\partial\beta}\nonumber\\
&=& K_{B}\ \beta^{2}\left(\left( \frac{A}{B}\right)^2-\frac{D}{B}\right).
\end{eqnarray}

Also, we calculate  Helmholtz's free energy $\ F(\beta)$ \cite{Hoover:2012csm} as:
\begin{equation}
\label{29}
F(\beta) = - K_{B}T\ln(Z(\beta)),
\end{equation}
leading to
\begin{equation}
\label{30}
F(\beta) = -\frac{1}{\beta}\ln\left( B \right).
\end{equation}
Then, from Eq. (\ref{29}) the entropy, $\ S(\beta)$, is concluded \cite{Hoover:2012csm} as,
\begin{equation}
\label{31}
S(\beta) = - \frac{\partial F(\beta)}{\partial T} = K_{B}\beta^{2}\frac{\partial F(\beta)}{\partial\beta},
\end{equation}
yielding (after some algebraic manipulations) the following expressions,
\begin{eqnarray}
S(\beta)= K_{B}\ {\beta} \left(  \frac{A}{ B}+\frac{\ln(B)}{{\beta}} \right)
\end{eqnarray}
Eventually, the chemical potential ${\mu(\beta)}$ calculated \cite{Hoover:2012csm} as,

\begin{eqnarray}
\label{35}
\mu(\beta)&=& -T\ \frac{\partial S(\beta)}{\partial N} = - \frac{1}{K_{B} \beta} \frac{\partial S(\beta)}{\partial N},\\ \nonumber
      &=&- \frac{1}{ N} \left(  \frac{A}{ B}+\frac{\ln(B)}{\beta} \right).
\end{eqnarray}

It should be noticed that one has to consider two different cases,
corresponding to high and low temperatures $T$ and compare with
$T_{rot}$ which determines the temperature scale relevant to rotational degrees of freedom. Figures~\ref{UFCS},~\ref{DeltaE}, and~\ref{a} show these thermodynamical quantities as a function of
$\beta$. All subsequent calculations have been performed for the parametrization of the potential given in Eq. (\ref{7}).
\begin{figure}[htpb]
\center
\includegraphics[width=0.75\textwidth]{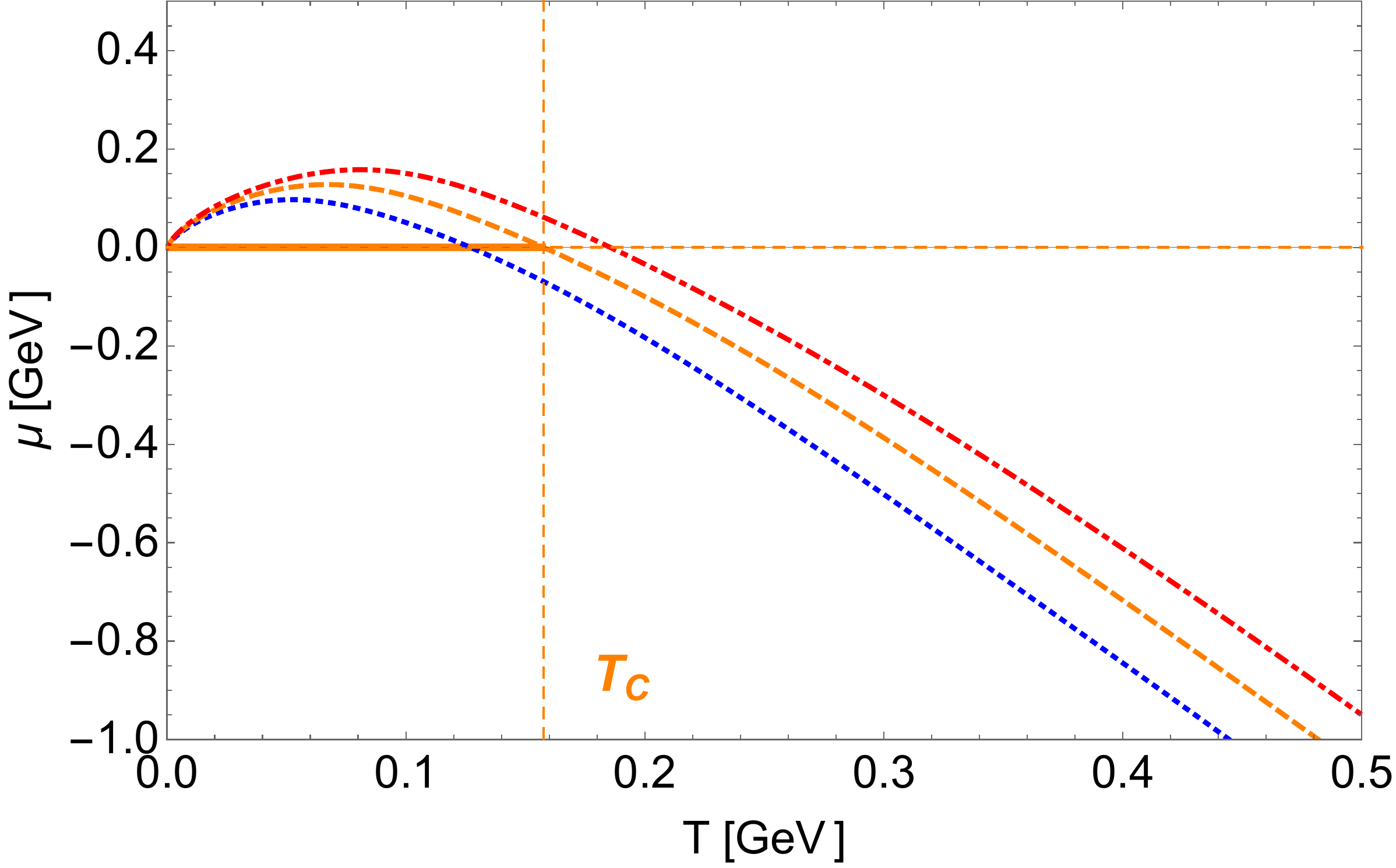} 
\caption{\label{b} The chemical potential as derived in Eq. (\ref{35}) as a function of temperature. $f_0(500)$ and charmonium values are denoted in red and orange colors, see the text for discussion.}
\end{figure}

\begin{figure}[htpb!]	
\center
\includegraphics[width=0.75\textwidth]{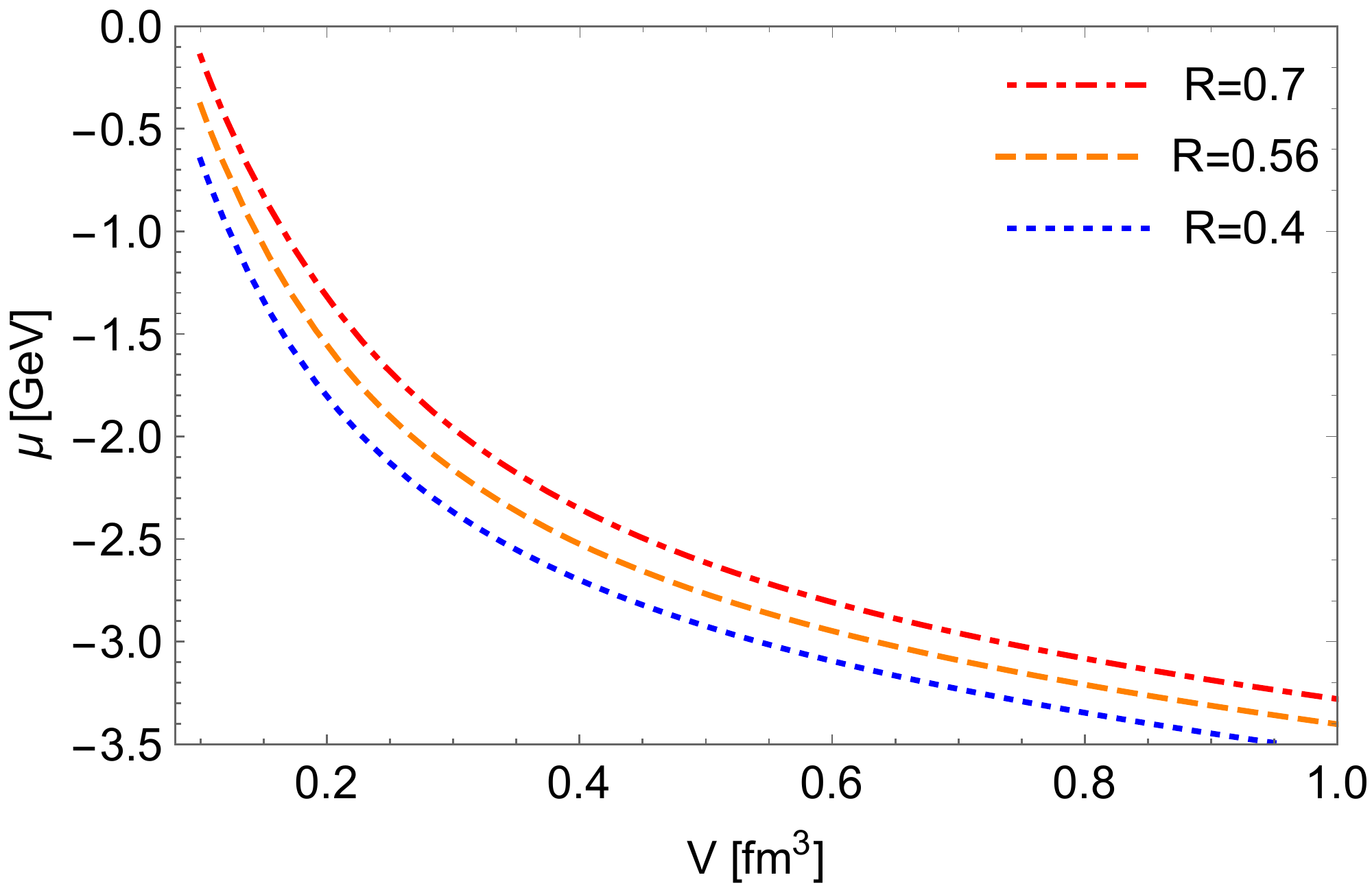}
\caption{\label{c} The chemical potential as derived in Eq. (\ref{35}) as a function of volume. $f_0(500)$ and charmonium values are denoted in red and orange colors.}
\end{figure}

Figure~\ref{b} shows the dependence of the chemical potential $\mu$ on the temperature $T$, whereas Figure \ref{c} shows the dependence of the chemical potential $\mu$ on the volume $V$. The bump of positive $\mu(T)$ values near origin appears as a typical artifact of using approximate methods in the evaluation of the thermodynamic properties of the Bose gas leading to expressions in closed form, an issue  discussed in~\cite{Sotnikov}. In reality, the $\mu(T)$ curve has to follow the plateau of vanishing values from origin to the critical point (thick segment on the horizontal axis) at which it bends and starts falling downward. Within this context, more elaborate numerical calculations would be required in future studies in order to rise the precision of the predictions of the critical points at which the phase transition takes place.

\subsection{Low temperature limit}

The low temperature limit,
  \begin{equation}
    T_{rot}>>T_c>>T_0=\frac{\hbar^{2} c^{2}}{mR^2},
    \label{insrt_1}
    \end{equation}
  equivalently,
  \begin{equation}
   1>>\frac{\hbar c}{mR}, \quad \mbox{or}\quad m>> \frac{\hbar c}{R},
   \label{insrt_2}
  \end{equation}
  where $m$ is the  mass of the particles in the gas, is of interest in thermal phenomena in QCD \cite{Hands}. Indeed, for
  \begin{equation}
    R<<\frac{\hbar c}{\Lambda_{QCD}}, \quad \Lambda_{QCD}\sim 200 \, \mbox{MeV},
    \label{insrt_3}
  \end{equation}
  with $\Lambda_{QCD}$ denoting the QCD scale parameter, it corresponds to the weak coupling regime of QCD, in which a perturbative description of phase transitions is possible. According to the $AdS_5/CFT_4$ gauge-gravity duality conjecture, such transitions appear dual to similar transitions in gravity, like the transitions of $AdS$ spaces to $AdS$ black holes. As a reminder, the $S^1\times S^3$ geometry emerges within this scheme  as the conformal boundary (the null ray cone) of the $AdS_5$ spacetime upon its conformal compactification.

In the case of low temperature, we may approximate the rotational partition function in Eq. (\ref{18}) by

\begin{equation}
\label{32}
 T \ll T_{rot} \rightarrow  Z(\beta)  \cong \frac{1}{2\gamma\beta}+4 e^{- 3\beta E_{0}}.
 \end{equation}
 The latter equation manifests the  exponential decrease of the partition function with  the increase of $\beta$. 
Then, the internal energy from Eqs. (\ref{23}) and (\ref{32}) calculates as:
\begin{equation}
\label{33}
 T \ll T_{rot} \rightarrow  U\left( \beta \right) =   \frac {A}{B},
 \end{equation}
 with
 \begin{eqnarray}
 A&=&{12 \ E_{0} \ e^{- 3\beta E_{0}}+\frac{1}{2 \gamma \beta^2}},
\nonumber\\
B&=&{4 \ e^{- 3\beta E_{0}}+\frac{1}{2\gamma\beta} }.
\end{eqnarray}
Moreover, from Eqs. (\ref{26}) and (\ref{32}) the variation of the energy (or "energy fluctuation") emerges as:
\begin{eqnarray}
\label{34}
T \ll T_{rot} \rightarrow  \left\langle {(\Delta U(\beta)) }^{2}\right\rangle &=& \left( \frac{A}{B}\right)^2-\frac{D}{B},
 \end{eqnarray}
with
 \begin{eqnarray}
 D&=&{36 \ E_{0}^2 \ e^{- 3\beta E_{0}}+\frac{1}{\gamma\beta^3}}.
\end{eqnarray}
In effect, the heat capacity  $C(\beta)$ is found as,
\begin{eqnarray}
 T \ll T_{rot} \rightarrow  C(\beta) &=&   - K_{B} \ \beta^2\left(\  \left( \frac{A}{B}\right)^2+\frac{D}{B}\right).
 \end{eqnarray}
Finally, from Eqs. (\ref{29}), (\ref{31}) and (\ref{32}),  Helmholtz's free energy and the entropy are calculated as:
\begin{equation}
\label{37}
 T \ll T_{rot} \rightarrow  F\left( \beta \right) = -\frac{1}{\beta}  \ln\left( B\right),
 \end{equation}
and
\begin{eqnarray}
\label{38}
T \ll T_{rot} \rightarrow  S(\beta) =  K_{B}\ \beta \left( \frac{A}{B}+\frac{\ln(B)}{\beta}\right),
 \end{eqnarray}
respectively. The low $T$ condition in (\ref{insrt_1}) is pretty well fulfilled by heavy flavor mesons, such as the charmonium, whose mass is $m_{c\bar c}=3097$ MeV. Indeed, for the corresponding hyperradius value of $R=0.56$ fm, used through the calculation, which has been fitted to the spectrum~\cite{Al-Jamel:2019myn}, the rotational temperature of $T_{rot}=352$ MeV is by about an order of magnitude larger than $T_0=40$ MeV.  This justifies hitting the Hagendorn's temperature as critical temperature for a Bose gas of such particles (vertical dashed line in Fig.~7).

  As to  the $f_0$ meson gas, the inequality in (\ref{insrt_1}) is fulfilled only for larger though not for much larger values between the involved quantities, which may explain why the related $T_c$ still falls in the reasonable range of changes around Hagedorn's critical temperature of $T_H=0.16$ GeV (c.f. \cite{Maiani}).

\subsection{High temperature limit}

In the high temperature limit the partition function is calculated as,
\begin{equation}
\label{41}
 T \gg T_{rot}  \rightarrow  Z(\beta) = \frac{\sqrt{\pi}}{4}\frac{1}{\left( E_{0}\beta \right)^{\frac{3}{2}}} e^{\beta E_{0}}.
 \end{equation}
Now, we turn our attention to the internal energy, given by
\begin{equation}
\label{42}
 T \gg T_{rot}  \rightarrow   U(\beta) = \ \left( \ \frac{3}{2\beta} - \ E_{0} \right) = \ K_{B}\left( \frac{3}{2}T - \ T_{rot} \right),
  \end{equation}
equivalently,  
\begin{eqnarray}
\label{43}
 T \gg T_{rot}  \rightarrow   U(\beta) &=& \frac{3}{2}\ K_{B}T \nonumber \\
						      &= & \frac{3}{2}\left(\frac{1}{\beta}\right).
\end{eqnarray}
The heat capacity $C(\beta)$ is evaluated as:
 \begin{equation}
\label{44}
 T \gg T_{rot}  \rightarrow   C(\beta)= \frac{3}{2}\left(\frac{1}{T \beta}\right)
  \end{equation}
Next we calculate the variation of the energy, Helmholtz's free energy, and the entropy as:
\begin{equation}
\label{45}
 T \gg T_{rot} \rightarrow  \left\langle {(\Delta U(\beta)) }^{2}\right\rangle=\frac{3}{2}\ {{(K}_{B}T)}^{2} = \ \ \frac{3}{2}\ {\left(\frac{1}{\beta}\right)}^{2},
  \end{equation}
 and,
 \begin{equation}
\label{46}
 T \gg T_{rot}  \rightarrow   F(\beta) = - \frac{1}{\beta}\ln\left( \frac{\sqrt{\pi}}{4}\ \left( E_{0}\beta \right)^{- \frac{3}{2}}
\  \right) - E_{0},
  \end{equation}
 and,
  \begin{equation}
\label{47}
 T \gg T_{rot}  \rightarrow   S(\beta) = {- K}_{B}\left[ \ \frac{3}{2} + \ln\left( \frac{\sqrt{\pi}}{4}\ \left( E_{0}\beta \right)^{- \frac{3}{2}} \right) \right].
  \end{equation}

\section{The grand canonical partition function $z(\beta,V,\mu)$ and related thermodynamic functions}
\label{sec:4}

The grand canonical partition function, here denoted by
$z (\beta,V,\mu)$ is related to the canonical partition function
$Z(\beta,R,b)$ for a system with N particles, as:
\begin{equation}
\label{48}
z\left(\beta,V,\mu \right) = \sum_{N = 0}^{\infty} \frac{1}{N!}\left(e^{\beta \mu } Z(\beta,R,b)\right)^N,
\end{equation}
where $\mu$ stands for the chemical potential energy.
For non-interacting systems where particles are free to move and exchange
positions (such as quantum gases), we know from the factorization
theorem \cite{Reif:2009fst} that the Eq. (\ref{48}) is reduced to:
\begin{equation}
\label{49}
z \left( \beta,V, \mu \right) = \frac{1}{\left(1-e^{\beta \mu} Z \left(\beta,R,b \right) \right)},  
\end{equation}
where V is the "volume" (three dimensional hypersurface) of the system which is well known \cite{Kirchbach:2016scz} and reads:
\begin{equation}
\label{50}
V = 2 \pi^{2}R^{3},
\end{equation}
implying,
\begin{equation}
\label{51}
R = \left( \frac{V}{2 \pi^{2}} \right)^{1/3}.
\end{equation}
As long as we need to study the canonical partition function
$Z(\beta,R,b)\ $as a function of the volume, we have to express $R$
in terms of$\ V$. We begin with Eq. (\ref{8}) with the aim to find the
volume dependence of the ground state energy$E_{0}$. In order to do
that, some detailed instructions regarding the calculations are needed.
We will continue taking  as an illustrative example of a color-charge dipole gas the
one constituted by $f_{0}\left( 500 \right)$ mesons, in which case the
parameters take the following values:
$\hbar c = 197.3289$ MeV fm, $Mc^{2} = 250$ MeV,
$N_{c} = 3$ and $R = 0.7$ fm. For the case of
 $f_{0}\left( 500 \right)$, the strong coupling constant
takes the value of $\alpha_{s} = 0.66 \pi $ \cite{Kirchbach:2016scz,Tanabashi:2018oca}. Substitution
of these values in Eq. (7), yields,
\begin{equation}
\label{52}
b = \frac{3 \times 0.66\ \pi\ }{2}.
\end{equation}
The above  potential parameters have been shown to fit~data on
the f$_{0}(500)$ meson excitation energies. Thus,
substituting above values and Eqs. (\ref{51}) and (\ref{52}) in to Eq. (\ref{8}), gives,
\begin{equation}
\label{53}
E_{0} = \frac{1.8}{V^{2/3}} , \qquad \gamma=\frac{0.56}{V^{2/3}}  \,\,  \mbox{[GeV]}.
\end{equation}
Hence, placing Eq. (\ref{53}) in to Eq. (\ref{Z_deg}), $Z\left( \beta,V,b \right)$ becomes,
\begin{equation}
\label{54}
Z\left( \beta,V,b \right) =\frac{0.8 \ V^{2/3}}{\beta}+4 \ e^{\left( \frac{-5.4\ \beta}{V^{2/3}} \right)}+ \frac{0.2 \ e^{\left( \frac{1.8\ \beta}{V^{2/3}} \right)}}{\left( \frac{\beta}{V^{2/3}} \right)^{3/2}}.
\end{equation}
Upon substitution of Eq. (\ref{54}) into Eq. (\ref{48}), $z\left( \beta,V,\mu \right)$ emerges as,
\begin{equation}
\label{55}
z\left( \beta,V,\mu \right) = \frac{1}{1- A}\nonumber,
\end{equation}
\begin{equation}
A=e^{\beta\mu}\left( \frac{0.8 \ V^{2/3}}{\beta}+4 \ e^{\left( \frac{-5.4\ \beta}{V^{2/3}} \right)}+ \frac{0.2 \ e^{\left( \frac{1.8\ \beta}{V^{2/3}} \right)}}{\left( \frac{\beta}{V^{2/3}} \right)^{3/2}}\right).
\end{equation}
Next we need to obtain the grand canonical potential \cite{Reif:2009fst} as,
{\small
\begin{eqnarray}
\label{56}
\Omega\left( \beta,V,\mu \right) &=& - K_{B} T \ln \left( z\left( \beta,V,\mu \right) \right)\nonumber \\
&=& \frac{1}{\beta} \ln \left(  1 -A \right).
\end{eqnarray} }

\begin{figure}[htpb!]	
\center
\includegraphics[width=0.75\textwidth]{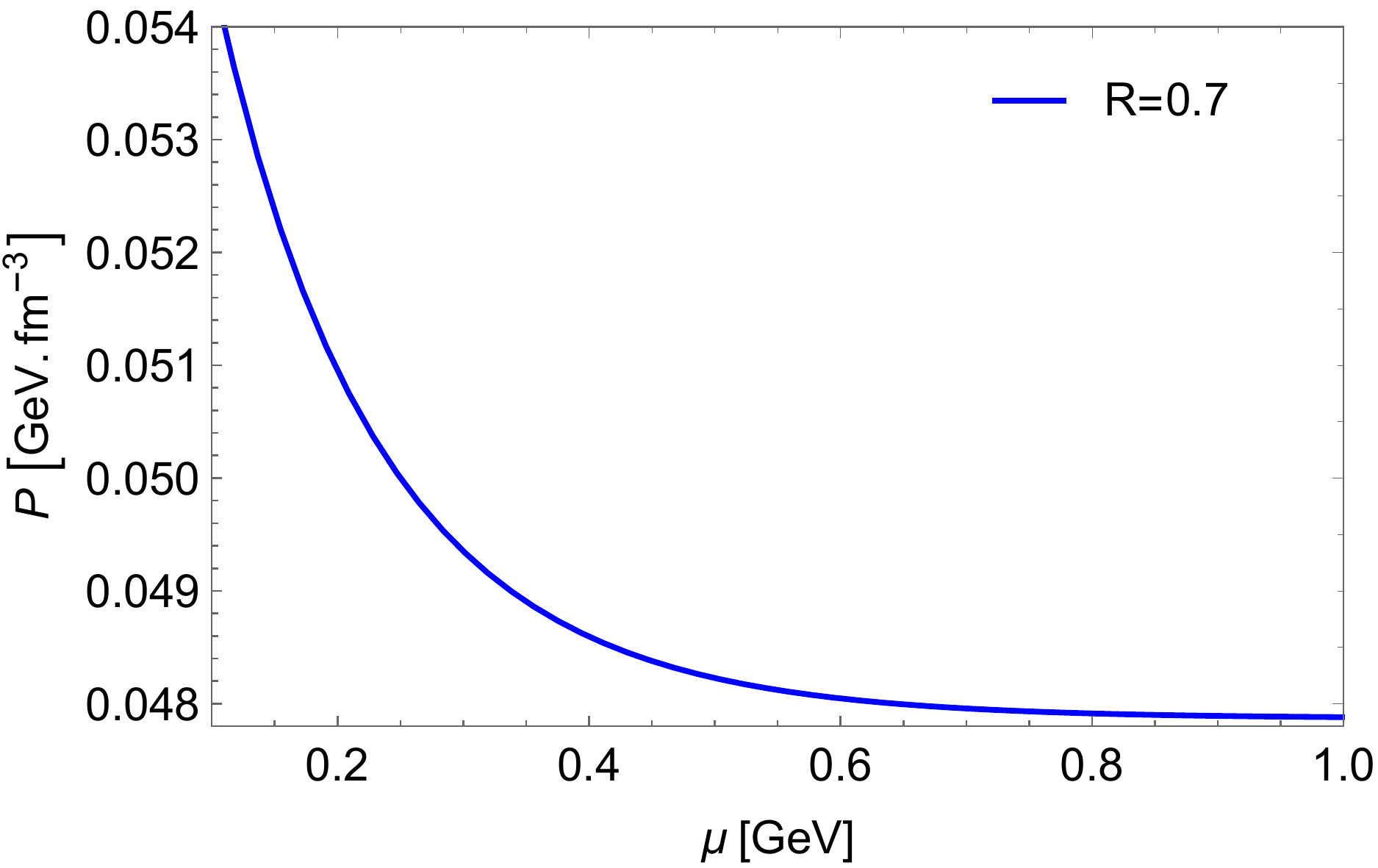}
\caption{\label{P-meson} The pressure as derived in Eq. (\ref{58}) as a function of ${\mu}$ for the $f_0(500)$ mesons. The parameters R and ${\alpha_s}$ take the value 0.7 fm and 0.66\ ${\pi}$, respectively.}
\end{figure}

\begin{figure}[htpb!]		
\center
\includegraphics[width=0.75\textwidth]{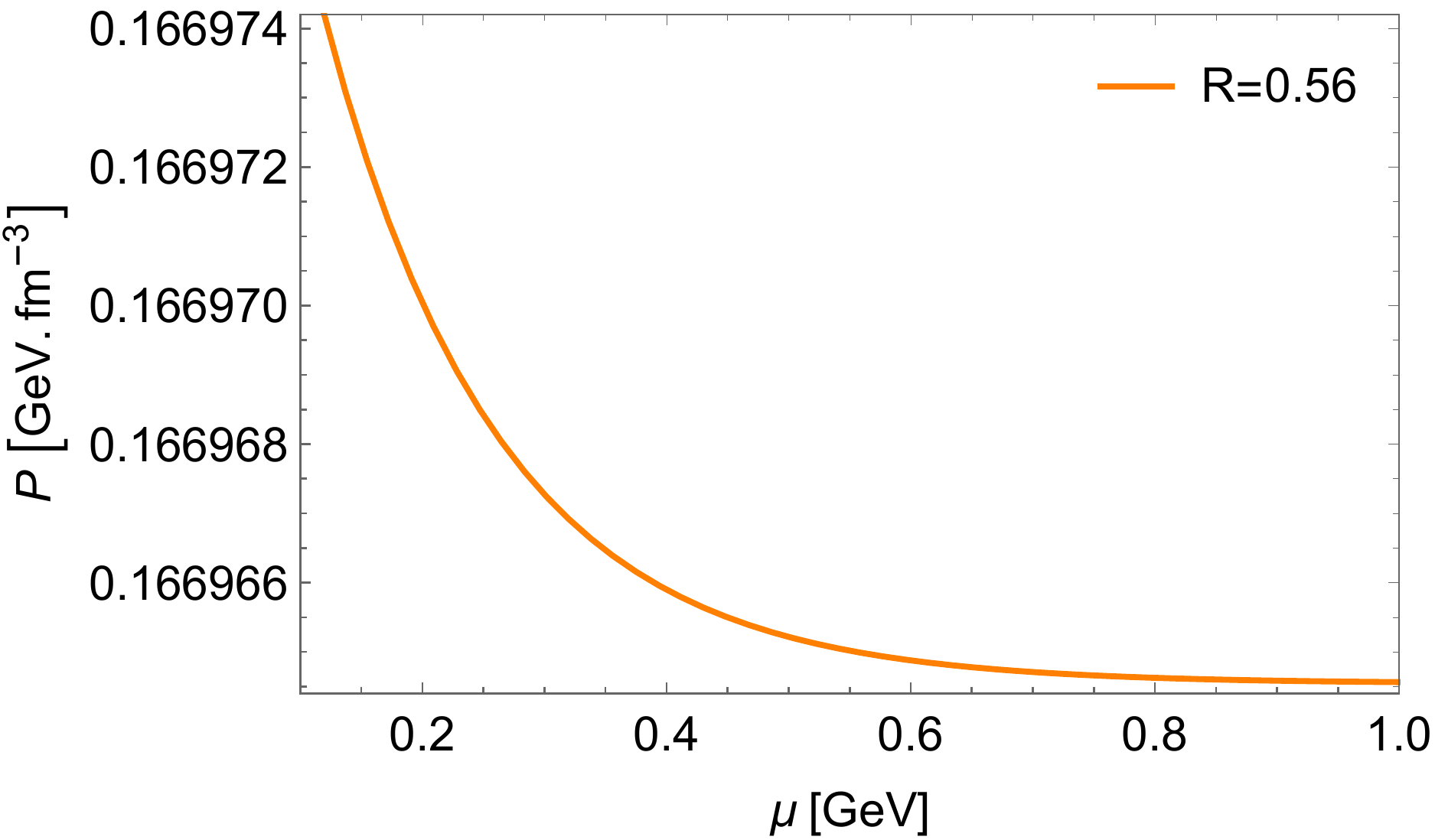}
\caption{\label{P-charmonium} The pressure as derived in Eq. (\ref{65}) as a function of ${\mu}$ for the charmonium. The parameters R and ${\alpha_s}$ take the value 0.56 fm and 0.2, respectively.}
\end{figure}

From the latter equation,  the particle number $N$ can be determined  as,
\begin{eqnarray}
\label{57}
N &=& - \left( \frac{\partial\Omega\left( \beta,V,\mu \right)}{\partial\mu} \right)_{\beta,V} \nonumber \\
&=& \frac{ A}{ 1 - A}.
\end{eqnarray}
Now one is ready to calculate  the pressure $P$ as,
\begin{eqnarray}
\label{58} 
P &=& -\left( \frac{\partial\Omega\left( \beta,V,\mu \right)\ }{\partial V} \right)_{\beta,\mu} = \frac{B\ e^{\beta\mu}}{\beta\left(1-A \right)}, \nonumber \\
B &= &\frac{0.56}{\beta\ V^{1/3}}+\frac{14\ \beta\ e^{\left( \frac{-5.4\ \beta}{V^{2/3}} \right)}}{V^{5/3}}+\frac{0.2\ \beta \ e^{\left( \frac{1.8\ \beta}{V^{2/3}} \right)}}{V^{5/3}\left( \frac{\beta}{V^{2/3}} \right)^{5/2}}
-\frac{0.2 \ \beta\ e^{\left( \frac{1.8\ \beta}{V^{2/3}} \right)}}{V^{5/3}\left( \frac{\beta}{V^{2/3}} \right)^{3/2}}.
\end{eqnarray}
\noindent
Finally, we are interested to study  the dependence of the pressure  on the strong coupling $\alpha_s$ and evaluate $P$ for the case of the
charmonium for which we adopt the $\alpha_{s} = 0.2$ value,
and set $Mc^{2} = 750$ MeV as in \cite{Tanabashi:2018oca,Al-Jamel:2019myn}. In so doing, the equation (\ref{8}) yields%
\begin{equation}
\label{59}
b = \frac{3 \times 0.2}{2}.
\end{equation}
Substitution of this value into Eq.(\ref{8}), yields
\begin{equation}
\label{60}
 E_{0} =  \frac{7.09}{V^{2/3}} , \qquad \gamma=\frac{0.2}{V^{2/3}}  \,\,  \mbox{[GeV]}.
\end{equation}
Hence, inserting  Eq. (60) into Eq. (41) allows us to write 
 $Z\left( \beta,V,b \right)$ as,
\begin{equation}
\label{61}
Z\left( \beta,V,b \right) = \frac{2.6 \ V^{2/3}}{\beta}+4 \ e^{\left( \frac{-21\ \beta}{V^{2/3}} \right)}+ \frac{0.02 \ e^{\left( \frac{7.09\ \beta}{V^{2/3}} \right)}}{\left( \frac{\beta}{V^{2/3}} \right)^{3/2}}.
\end{equation}
Now substitution of Eq. (\ref{61}) into Eq. (\ref{49}) leads to
\begin{eqnarray}
\label{62}
z\left( \beta,V,\mu \right) &=& \frac{1}{1- A },\nonumber\\
A&=&e^{\beta\mu}\left( \frac{2.6 \ V^{2/3}}{\beta}+4  e^{\left( \frac{-21\ \beta}{V^{2/3}} \right)}+ \frac{0.02 \ e^{\left( \frac{7.09 \beta}{V^{2/3}} \right)}}{\left( \frac{\beta}{V^{2/3}} \right)^{3/2}}\right).
\end{eqnarray}
With that, the grand canonical potential $\Omega$  calculates as,
\begin{eqnarray}
\label{63}
\Omega\left( \beta,V,\mu \right) = \frac{1}{\beta}\ln\left(1-A\right) .
\end{eqnarray}
and allows to  express  the number of particles  $N$ according to,
\begin{equation}
\label{64}
N = \frac{A}{1-A}.
\end{equation}
In effect,  the pressure $P$ takes the form of,
\begin{equation}
\label{65}
P = -\frac{e^{\beta\mu}F}{\beta D},
\end{equation}
with
\begin{eqnarray}
F&=&\frac{1.75}{\beta\ V^{1/3}}+\frac{56\ \beta\ e^{\left( \frac{-21\ \beta}{V^{2/3}} \right)}}{V^{5/3}}+\frac{0.02\ \beta \ e^{\left( \frac{7.09\ \beta}{V^{2/3}} \right)}}{V^{5/3}\left( \frac{\beta}{V^{2/3}} \right)^{5/2}}-\frac{0.11 \ \beta\ e^{\left( \frac{7.09\ \beta}{V^{2/3}} \right)}}{V^{5/3}\left( \frac{\beta}{V^{2/3}} \right)^{3/2}}, \nonumber\\
D&=& \beta\left({ 1 -  H\ e^{\beta\mu}}\right),\nonumber\\
H&=&\left( \frac{2.6 \ V^{2/3}}{\beta}+4 \ e^{\left( \frac{-21\ \beta}{V^{2/3}} \right)}+ \frac{0.02 \ e^{\left( \frac{7.09\ \beta}{V^{2/3}} \right)}}{\left( \frac{\beta}{V^{2/3}} \right)^{3/2}}\right).\nonumber
\end{eqnarray}
The pressure as a  function of the chemical potential is plotted in the Figs.~\ref{P-meson} and \ref{P-charmonium}, where the parameter $\beta$ takes the value of 4 GeV$^{-1}$. Compared to the pressure in  the $f_{0}(500)$ meson gas (shown  in Fig.~\ref{P-meson}) the pressure for the charmonium (illustrated by Fig.~\ref{P-charmonium})  
presents itself  significantly higher. This means that the charmonium gases 
are more compact than $f_0(500)$ meson systems,  as it should be.

\section{Summary and Conclusions}

In this work we studied the thermodynamic properties of the
trigonometric Rosen-Morse potential in the parametrization of Eqs. (\ref{3})
and (\ref{7}), and from the perspective of its utility as an effective
potential in color-neutral quark systems. The parametrization chosen
allowed to cast the one-dimensional Schr\"odinger potential problem as a
perturbation of free quantum motion on the three-dimensional
hyper-sphere, $S^{3}$, a manifold of finite volume that can host only charge-neutral
systems, such as color-charge neutral mesons. It has to be noticed that
preferring the hyper-sphere is not essential for the modelling of the
charge-neutrality because any manifold without boundary, be it symmetric
or not, can bear only charge-neutral systems. The $S^{3}$ manifold  brings the
advantage of a particularly simple form of the Laplace-Beltrami operator
that gives rise to an exactly solvable dynamics and degeneracy patterns
in the excited levels of the type observed in some light mesons, such as
$f_{0}$ and $a_{0}$ \cite{Kirchbach:2016scz}. The potential under discussion has
found applications besides in the sector of the light unflavored mesons
\cite{Kirchbach:2016scz}, also in heavy quarkonia \cite{Al-Jamel:2019myn}. In both cases it has been
found to provide quite a realistic data description. Therefore, the
canonical and grand canonical partition functions calculated here
account for 
\begin{itemize}
\item color electric charge neutrality,  
\item multiplicities of states in the levels,
\item finite volumes.
\end{itemize}

Accounting for the state multiplicity  effects, caused an increase in the values of the Canonical partition function. A similar consequence is observed due to the shrinkage of the volume of the system though this effect is not that significant in magnitude  as the presence of  degeneracy. Our results were illustrated for the case of the charmonium and the $f_0(500)$ meson. In particular we observed that a charmonium gas is more compact as the one constituted by $f_0(500)$ mesons, as is to be expected.From  Fig. 7  we read off quite reasonable values of the critical temperature as  $T_c\in [0.157 -0.175]$  GeV, a range of variations around Hagedorn's value also quoted  by other authors \cite{Maiani}.

To the amount the majority of the mesons detected so far are commonly
accepted to be color-anti-color (quark-anti-quark) dipoles, a picture
also adaptable to baryons when considered as quark-diquark
configurations, therefore we expect the finding of the present study to be useful
to studies of thermodynamic aspects of QCD phenomena~\cite{Ayala:2007cp}.
The advantage of the method presented here is that in being based on fundamental color dipoles, it allows, by the aid of the techniques known from the physics of diatomic molecules, to track down phase transitions from gaseous to liquid states of colorless quark matter and considering the dipole dissociation \cite{Dubinin:2013yga} towards colored states.

Furthermore, this approach allows to control the state of matter through the topology of the internal space \cite{Casetti:2000gd} of hadrons whose running curvature could be made temperature dependent and allowed to vary between positive and negative values. The latter case would not prohibit observability of single charges and is expected to provide a scenario toward  color deconfinement. Consequently, this manuscript provides the necessary building blocks required by future phase transition studies \cite{Elliott:1999fj} within a quantum mechanical approach to QCD, which manifestly takes into account the fundamental field theoretical concept of colorless-ness of hadrons and which in addition explicitly depends on the strong coupling constant and the number of colors.

All in all, we have worked out  the canonical and grand canonical partition functions related to the trigonometric Rosen-Morse potential together with the principal thermodynamic functions and consider them as promising tools  for an adequate description of color neutral states of hadronic matter as they appear in heavy-ion collisions and applications to matter in compact stars, a topic for future research.

 \section{Acknowledgements}

We thank David Blaschke for inspiring discussions on the aspects of quark confinement. D. A-C. acknowledges support from the the Bogoliubov-Infeld program for collaboration between JINR and Polish Institutions as well as from the COST actions CA15213 (THOR) and CA16214 (PHAROS).

\end{document}